# pH-Switchable Pickering Emulsions Stabilized by Polyelectrolyte-Biosurfactant Complex Coacervate Colloids

Sandrine Laquerbe,[a] Alain Carvalho,[b] Marc Schmutz,[b] Alexandre Poirier,[a] Niki Baccile,[a,*] Ghazi Ben Messaoud,[a]†,*

[a] Sorbonne Université, Centre National de la Recherche Scientifique, Laboratoire de Chimie de la Matière Condensée de Paris, LCMCP, F-75005 Paris, France

[b] Université de Strasbourg, CNRS, Institut Charles Sadron UPR 22, 67034 Strasbourg, France

† Current address: DWI- Leibniz Institute for Interactive Materials, Forckenbeckstrasse 50, 52056, Aachen, Germany

* Corresponding author:

E-mail address: niki.baccile@upmc.fr; benmessaoud@dwi.rwth-aachen.de

**TOC**

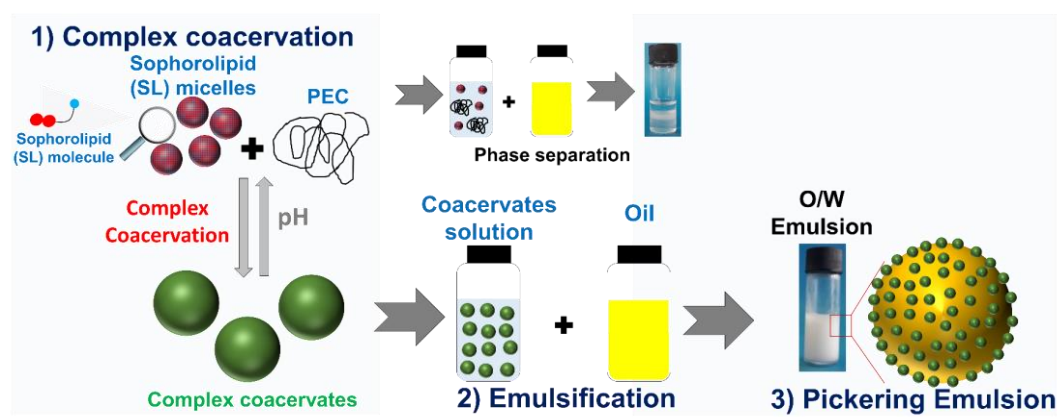

## Abstract

**Hypothesis**

Polyelectrolyte-surfactant complexes (PESCs) have long been employed as oil-in-water (o/w) emulsions stabilizers, but never in the structure of colloidal complex coacervates providing a Pickering effect. The complexed state of PESCs could make them unsuitable o/w Pickering emulsifiers, which instead require a balance between colloidal structure and stability, amphiphilicity and wettability. Here we hypothesize that PESCs coacervates are efficient Pickering stabilizers. Instead of classical surfactants, we employ sophorolipid (SL) biosurfactants, atypical anionic/neutral stimuli-responsive biosurfactants. Despite their tunable charge and mild amphiphilic character, they can be used in combination with cationic/neutral polyelectrolytes (chitosan, CHL, or poly-L-lysine, PLL) to form PESC coacervates for the development of biobased, but also pH-switchable, Pickering emulsions.

**Experiments**

Aqueous solutions of SL-CHL (or SL-PLL) complex coacervates are emulsified with dodecane. Confocal laser scanning microscopy (CLSM) and scanning electron microscopy under cryogenic conditions (cryo-SEM) demonstrate the Pickering effect, while optical microscopy and oscillatory rheology respectively assess the emulsion formation and relative viscoelastic properties.

**Findings**

Both SL-CHL and SL-PLL PESCs stabilize o/w emulsions up to $\Phi_{oil}$ of 0.7 only in the pH region of complex coacervation ($6 < pH < 9$): outside this range, phase separation occurs. Rheology shows a typical solid-like response and mechanical recovery upon applying large deformations. CLSM and cryo-SEM highlight a colloidal structure, associated to the complex coacervates, of the oil/water interface and suggest a Pickering effect. These findings demonstrate the Pickering effect from PESC coacervates and the possibility to use biobased and biocompatible components, with application potential in cosmetics, food science, or oil recovery.

***Keywords:*** *Surfactants; Polyelectrolytes; Polyelectrolyte Surfactant Complex (PESC); Bolaamphiphiles; Sophorolipids; Complex coacervates; Pickering emulsion; pH-responsive.*

# 1. Introduction

Emulsions, mixtures of two immiscible liquids, are used for a wide range of applications [1,2]. Emulsions, thermodynamically unstable, are traditionally stabilized using surfactants, polysaccharides, proteins, or complexes.[3] However, besides the traditional use of surfactants, emulsion stabilization can also be achieved by solid particles. Pickering emulsion, first discovered by Ramsden[4] and Pickering,[5] refers to particle-stabilized interfaces; they are assumed to be more stable against coalescence than surfactant-stabilized emulsions due to the high energy required for particle detachment from the droplet surface.[6] Historically, particles are composed of organic or inorganic solids of natural and synthetic origin [7]. Still, more than a decade ago, soft particles like cross-linked polymer microgels were also tested.[8] Unlike solid particles that are partially wetted by water and oil and firmly anchored at the oil-water interface, soft microgels are highly swollen by the water phase. Their deformability is the main factor controlling emulsion stability.[9] The use of soft colloids to stabilize interfaces has brought new functionality to Pickering emulsions, otherwise too stable and requiring a high energetic input to break.[10] On-demand control of the emulsion stability is of great importance for applications like oil recovery,[11] liquid phase heterogeneous catalysis,[12] and emulsion polymerization,[13] but also towards sustainable chemistry by recycling both aqueous and organic phases in view of reuse of the emulsifier.[10] The development of stimuli-responsive emulsions has consequently drawn much attention due to their reversible nature.[10]

The quest for soft, stimuli-responsive and reversible, Pickering stabilizers has recently shown the possibility to use an old family of soft colloids, like oppositely charged inter-macromolecule complexes, generally referred to as complex coacervates, [14] and generally containing polyelectrolytes (PEC) and/or proteins.[15,16] Nonetheless, to date, only a limited number of studies, mainly involving PEC-PEC or polysaccharide-protein complex coacervates, were employed as emulsion stabilizers via a Pickering process [17–23]. If these studies demonstrate the interest of complex coacervates colloids as emulsion stabilizers, they also introduce interesting questions in the broader field of colloids science, namely the amphiphilic character of complex coacervates[24] and their stability at complex interfaces, hence, the actual mechanism behind the Pickering structure.

The same questions are even more intriguing when another class of soft complexes are employed as emulsion stabilizers, polyelectrolyte-surfactant complexes (PESC) [25–27].

These systems have been long studied at oil-water interfaces, [28] but never in the shape and structure of complex coacervate colloids forming a Pickering emulsion. Simple questions like the stability itself of PESC coacervates at the oil-water interface, hence the possibility to form PESC-based Pickering emulsions, but also more complex questions like their wettability, composition or deformability are simply unanswered.

Instead of classical surfactants, like alkyl sulfates or alkyl quaternary ammonium salts, we employ biosurfactants, because of their environmental and physicochemical interest. [29,30] Considering the more stringent regulations combined with the worldwide consumers' concerns for health and environment, biosurfactants are the obvious eco-friendly option to develop potentially commercial alternatives to petrochemical surfactants. In the meanwhile, biosurfactants constitute a new category of functional bolaform amphiphiles, of which the micellar shape, charge, aggregation number and amphiphilic character change with pH.[31] In this regard, the colloidal behavior of biosurfactant-based PESC coacervates as Pickering emulsion stabilizers is even riskier and intriguing than classical surfactants.

Sophorolipids are an attractive class of biobased surfactants exclusively produced from renewable agro-resources through a fermentation process of the yeast *S. bombicola*, with interesting production rates and reduced environmental impact biosynthesis. [32] Sophorolipids possess several attractive applications in cosmetics[33] or as antimicrobial agents[34] or soft materials preparation.[35] In this regard, sophorolipids are known to have surface-active properties and they are suitable for stabilizing oil-in-water emulsions.[36]

Compared to conventional surfactants, there have been only a few studies of sophorolipids' emulsifying ability. Koh *et al*. have shown that sophorolipids' interfacial performance is best achieved in a formulation composed of the lactonic and acidic forms,[37–40] rather than each individual congener. However, the interfacial performance remains relatively low.[38] The poor emulsifying performances were attributed to the bolaform, double hydrophilic, nature of acidic sophorolipids, and to the poor surface-active character of lactonic sophorolipids. The surface charge of sophorolipid micelles is pH-dependent[41] and the micellar structure itself does not have a well-defined hydrophilic-hydrophobic character.[31] For these reasons, they developed a series of sophorolipid-ester derivatives, with enhanced emulsifying performances [37–40] on paraffin (alkane-based), almond (triglycerides-based), or lemon (cyclic and partially hydrophilic terpenes) oils. As a function of the oil phase's nature, sophorolipid concentration, and chemical structure, an oil-in-water (o/w) emulsion containing up to 0.1 oil volume fraction are formed and are stable during few days.[37–39]

We have recently reported the complex coacervation between sophorolipid micelles and cationic PEC.[42] Interestingly, complex coacervation is driven by pH by tuning the electrostatic interaction strength between SL and PEC, where only deacetylated acidic sophorolipids are used. In the current study, we employ pH-responsive sophorolipids-PEC complex coacervates to form reversible o/w emulsions. Similarly, sophorolipids are employed in their deacetylated open acidic form without any chemical modification. We show that emulsion stabilization occurs by a Pickering effect via the pH-responsive complex coacervates adsorption at the oil-water interface (oil = dodecane). We show that the o/w Pickering emulsions are stable up to $\Phi_{oil}$ of 0.7 and that emulsion can be easily broken and reformed by a simple pH switch method. The emulsion droplets' size distribution is examined using optical microscopy, while the solid-like behavior is probed by rheology. The colloidal stabilization of the interface is eventually demonstrated by combining confocal laser scanning microscopy and scanning electron microscopy under cryogenic conditions, both avoiding dehydration and preserving the emulsion structure.

## 2. Experimental Section

### 2.1. Chemicals

Sophorolipids ($M_w$= 622 g.mol$^{-1}$) are purchased from Soliance (Givaudan Active Beauty, France) and hydrolyzed in an alkaline medium. The deacetylated open acidic form is obtained by lowering pH at around 4.5 and finally recovered using the standard pentanol method.[43] The final compound is constituted of deacetylated (open acidic) sophorolipids (SL) with a C18:1 (subterminal ω-1) content above 80% and forming a stable micellar phase in a broad acidic-basic pH range.[44] Chitosan oligosaccharide lactate (CHL) ($M_n$≈ 5 KDa, $pK_a$≈ 6.5, deacetylation degree >90%), poly-L-lysine hydrobromide (PLL) ($M_w$≈ 1-5 KDa, $pK_a$≈ 10) and FITC-labeled poly-L-lysine ($M_w$≈ 15-30 KDa, 0.003-0.01 moles FITC per mole of lysine monomer) are purchased from Sigma-Aldrich. 18:1 Liss Rhod PE (1,2-dioleoyl-sn-glycero-3-phosphoethanolamine-N-(lissamine rhodamine B sulfonyl) (ammonium salt), $M_w$= 1.3 KDa) is purchased from Avanti Lipids®. This compound is labeled Rho-PE. Dodecane (Reagent

Plus, purity≥ 99%) from Sigma-Aldrich is used to prepare emulsions. All other chemicals are used without further purification.

## 2.2. Preparation of complex coacervate solutions

Complex coacervates are prepared according to our previous study.[42] Briefly, SL and PEC's stock solutions are prepared separately by dissolving each compound's appropriate amount in Milli-Q-grade water. Then, SL solution and a selected PEC solution are mixed together with final concentrations (C) of $C_{SL}$= 5 mg/mL, $C_{CHL}$= 1.4 mg/mL and $C_{PLL}$= 2 mg/mL. The mixture's final pH is then adjusted under mild stirring by adding a few µL of HCl (0.1 M or 0.5 M) or NaOH (0.1 M or 0.5 M).

## 2.3. Preparation of emulsions

A known volume fraction of dodecane, $\Phi_{oil}$ ($0 \leq \Phi_{oil} \leq 0.75$) is mixed with (1- $\Phi_{oil}$) of a SL-PEC complex coacervate solution to prepare a biphasic system. The mixture is then subjected to a high-speed homogenization (rotor-stator) using an Ultra-Turrax homogenizer (IKA® T25 Basic Werke) with a dispersing element of 25 mm (stator diameter) to obtain a homogeneous emulsion. Homogenization is maintained for 1 min at a constant speed of 13500 rpm with the vial's vertical movements along the rod.

## 2.4. Pendant drop tensiometry

The drop shape analysis system DSA30 Krüss, Germany, is used with associated software and microsyringes SY20 of 1 mL in borosilicate glass. The cleanliness of the setup is verified by pumping 10 times the syringe volume with milliQ water. The surface tension must be constant and reproducible ±0.5 mN/m during the total time of the experiment. A pendant drop of 11 – 30 µL of the solution is produced in air with a steel capillary having an external diameter of 1.83 mm. Images are recorded each 1 s during 300 s. Contour of the drop is fitted by the Young-Laplace equation using an iterative process with the surface tension, $\sigma$, as an adjustable parameter.

## 2.5. Optical microscopy

To visualize the emulsion droplets, images are acquired using Nikon DS-Ri1 optical microscope in brightfield mode. The emulsion droplets' size distribution, after preparation and following destabilization/re-emulsification, is estimated by image analysis using Fiji software (National Institutes of Health, Bethesda, Mayland, USA) to calculate diameters of about 100 droplets for each sample.[45]

## 2.6. Rheology

The o/w emulsions' rheological properties are carried out at 20°C using an MCR 302 rheometer (Anton Paar, Austria) equipped with a solvent trap to ensure minimal water evaporation during the measurements. The oscillatory shear experiments are performed using a sandblasted plate-and-plate geometry (25 mm) with an initial gap of 0.5 mm and a controlled normal force (NF ~ 0 N). First, a dynamic strain sweep is conducted at an angular frequency ($\omega$) of 6.28 rad/s by varying the shear strain ($\gamma$) from 0.01 to 1000 % to determine the linear viscoelastic region. A value of strain within the linear viscoelastic regime is then applied in the following angular frequency sweep between 100 and 0.01 rad/s.

Step strain experiments under small and large strains are performed to evaluate the mechanical properties' potential recovery. For each emulsion, five cycles of destructuring/restructuring are performed. The sample is subjected to a large strain (20% for PLL emulsion) for 5 minutes, then to a small strain (0.3%) for 30 minutes (except during the last cycle where the small strain lasts 1 hour). All measurements are carried out 24 hours after emulsions preparation.

## 2.7. Confocal Laser Scanning Microscopy (CLSM)

The obtained emulsions are observed using a Leica TCS SP5 (Leica Microsystems, Heidelberg, Germany) equipped with an internal GaAsP hybrid detection system. To visualize the complex coacervates, two sets of experiments are performed using solely FITC-labeled PLL or both FITC-labeled PLL and a rhodamine-labeled C18:1 lipid (Rho-PE). The samples are excited with FITC-PLL and Liss Rhod PE specific excitation wavelengths of 488 nm and 561 nm. The emission is detected between 498-548 nm and 571- 631 nm, respectively, using photomultiplier tube (PMT) detectors. CLSM images are analyzed using Fiji software (National Institutes of Health, Bethesda, Mayland, USA) to acquire diameters of about 100 droplets for each sample.[45] To avoid any significant effect of FITC-PLL or Rho-PE on the complex coacervate and emulsion formation, several PLL:FITC-PLL mass ratio and SL:Rho-PE molar ratios are priory tested, and a mass ratio of 57 and a molar ratio of 500 are selected, respectively. The experimental conditions for complex coacervates labeling are given in Table 1.

Table 1 – Concentration values used to prepare labeled SL-PLL complex coacervate solutions. $\frac{m_x}{m_y}$ and $\frac{mol_x}{mol_y}$ respectively refer to mass and molar ratio between compounds x and y. All concentrations are intended in milli-Q-grade water before mixing with the oil phase. For both samples, stock solutions of FITC-PLL ($C_{FITC-PLL}$= 1.43 mg/mL in water) and Rho-PE ($C_{Rho-PE}$= 2.69 mg/mL in absolute ethanol) are preventively prepared and stored at -18°C.

| Experiment | $C_{SL}$ / mg/mL | [SL] / mM | $C_{PLL}$ / mg/mL | $C_{FITC-PLL}$ / mg/mL | $C_{Rho-PE}$ / mg/mL | [Rho-PE] / mM | $\frac{m_{PLL}}{m_{FITC-PLL}}$ | $\frac{mol_{SL}}{mol_{Rho-PE}}$ | pH |
|---|---|---|---|---|---|---|---|---|---|
| Single labeling | 5 | 8.1 | 1.9 | 0.024 | - | - | 76 | - | 8.43 |
| Double labeling | 5 | 8.1 | 1.8 | 0.025 | 0.021 | 0.016 | 76 | 501 | 6.26 |

## 2.8. Scanning Electron Microscopy under cryogenic conditions (cryo-SEM)

A small volume of a selected emulsion is plunged into slush nitrogen. The frozen sample is transferred into the Quorum PT 3010 chamber attached to the microscope. There, the frozen sample is coated with a thin Pt layer (by sputter deposition) and fractured with a razor blade. A slight etching of the sample is performed at -90 °C for 2 minutes. No further metallization step is required before transferring the sample to the SEM chamber. The adsorption of coacervates onto oil droplets is observed using a cryo FE-SEM (Hitachi SU8010) imaging at -150 °C and a voltage of 1 keV.

## 3. Results and discussion

### 3.1. Emulsion formation

In a recent publication, we have shown that SL micelles form pH-responsive complex coacervates in the presence of PEC, such as chitosan oligosaccharide (CHL) or poly-L-lysine (PLL).[42] If PLL is not known to be involved in emulsion stabilization, both the basic form of chitosan alone [46,47] and its complexes with surfactants (although not in a complex coacervate phase) [48–50] have been proven to be suitable emulsion stabilizers. Our previous work, summarized in Figure 1 and Figure S 1 commented in this paragraph, showed four distinct regions, the pH range of which depends on the respective pKa of SL and PEC, in the titration curves of both SL-CHL (Figure 1) and SL-PLL (Figure S 1). In region 1, *off-plateau*, the solution is clear, and turbidity is close to zero, SL forms uncharged micelles, and PEC is generally soluble.[31,41,51] In region 2, turbidity progressively increases (the solution becomes cloudy), SL micelles start to become negatively charged due to the carboxylic acid's deprotonation, thus promoting electrostatic interactions with the positively charged PEC. In region 3, between the pKa's of SL and PEC corresponding to the *on-plateau*, turbidity is maximized; SL and PEC form stable complex coacervates. In Region 4, *off-plateau*, turbidity decreases again, the solution becomes clear, SL forms negatively-charged micelles[31,41,51] and the PEC is either soluble (PLL) or insoluble (CHL). The coacervation process is optimized on the plateau (*on-plateau*) region of the titration curve, which is between approximately pH= 5.5 and 6.9 for CHL (Figure 1) and pH= 5.7 and 8.5 for PLL (Figure S 1).

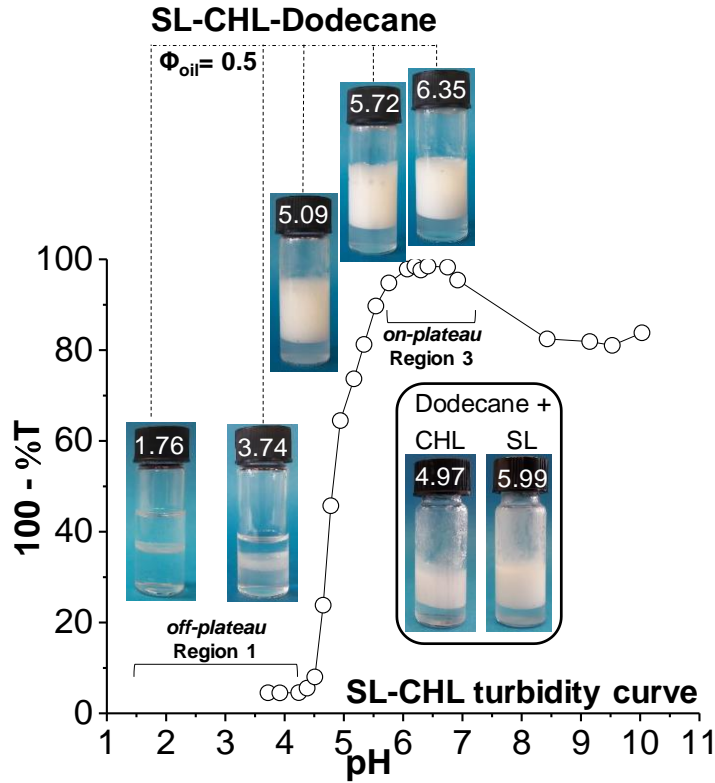

**Figure 1** - Turbidity profile (100%-T, empty circles) as a function of pH for a SL-CHL mixture ($C_{SL}$= 5 mg/mL; $C_{CHL}$= 1.4 mg/mL, data reproduced from Ben Messaoud et al.,[42] Complex coacervation is maximized on-plateau, here between pH 5.5 and 7. Each vial along the pH curve contains o/w emulsion with a $\Phi_{oil}$ of 0.5 and an SL-CHL solution prepared at different pH-values(indicated on top of each vial). The two control vials in the black box refer to o/CHL-(left-hand side, SL-free) and o/SL-(right-hand side, CHL-free) emulsions respectively prepared at pH= 4.97 and pH= 5.99 ($\Phi$oil= 0.5).

The surface-active properties of SL, CHL, PLL and their corresponding complex coacervates prepared *on-plateau* are compared to water in Figure 2 a) and b) through dynamic surface tension experiments. As expected, water's surface tension is constant, close to 72 mN/m, during the entire experimental time. Both CHL and PLL have a time-independent evolution of the surface tension, similar to water, thus displaying no surface-active properties, except for an abrupt variation of $\sigma$ in the case of PLL above 100 s and which we attribute to spurious impurities in the PLL solution. SL solutions prepared at both pH 5.7 and 6.3 are clear, and they induce an immediate drop of the surface tension from 72 to 40 mN/m at $t = 0$ s. The adsorption/desorption equilibrium is obtained after 15 – 50 s, with $\sigma = 32 \pm 0.5$ mN/m. SL-PLL and SL-CHL solutions, respectively prepared *on-plateau* at pH 6.3 and 5.7, show the classical turbidity expected for surfactant-polyelectrolyte complex coacervates in solution. They also display a drop in surface tension at t= 0 s, comparable to the free SL solution, thus suggesting that the non-complexed, free, SL component is firstly adsorbed at the air-water interface. However, the surface tension evolves more slowly with time, and no well-defined adsorption/desorption equilibrium is reached after 300 s for the complex coacervates solutions compared to the SL one.

Interestingly, water evaporation drives the formation of an opaque, thick crust after t= 1h20 at the drop's surface containing the SL-CHL solution. After filling the drop with a sequential increment of the filling solution (images f)-g) in Figure 2) shows the formation of a pending drop having a non-Laplacian shape. A Laplacian geometry of the drop is commonly observed for low-molecular-weight surfactant solutions due to adsorption/desorption equilibrium, which produces a fluid film. Laplacian-shaped drops are also observed here for pure water, SL and CHL solutions (images c)-e), Figure 2), respectively having σ= 72, 40 and 32 mN/m.

On the contrary, high molecular weight surfactants, proteins, and other amphiphilic polymers irreversibly adsorb at the air/water interface, and they produce thick multi-layer films. Complex coacervate solutions studied here also seem to exhibit an irreversible interface adsorption, producing a thick layer, which deforms the shape of the drop. For this reason, despite the comparable value of surface tension after t = 300 s between SL and SL-PEC solutions, we can safely state that the evolution of σ in the complex coacervates solutions is driven by a combined surface tension effect attributed to both free SL (fast) and SL-PEC complexes (slow). In light of the above, we test SL-PEC solutions' ability to stabilize o/w emulsions according to the on-plateau (Region 3) or off-plateau (Region 1) positioning on the SL-PEC turbidity curve.

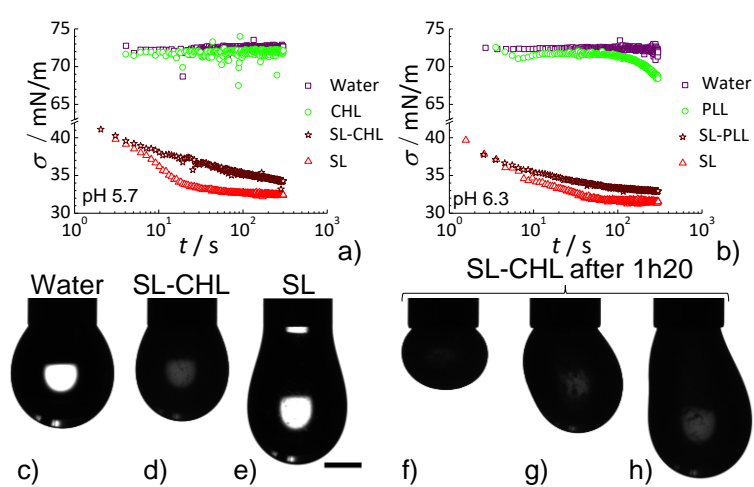

**Figure 2 - a, b) Dynamic surface tension experiments using the pending drop apparatus of water. Composition of the solutions: water (mQ-grade water, no pH adjustment), CHL (1.4 mg/mL, pH 5.7), PLL (2 mg/mL, pH 6.3), SL (5.0 mg/mL, pH 5.7 in a) and pH 6.3 in b)), SL-CHL (respectively 5.0 and 1.4 mg/mL, pH 5.7), SL-PLL (respectively 5.0 and 2 mg/mL, pH 6.3). c-e) Images of characteristic pendant drops (12 ± 1 µL) composed of c) water 72 mN/m, d) SL-**

**CHL 40 mN/m and e) SL 32 mN/m solutions. f-h) Images of SL-CHL drops after 1h20. For each picture few µL of solution is added. The scale bar (same for all images) in e) corresponds to 1 mm.**

Based on these results, a SL-CHL solution is mixed with an isovolumetric equivalent of dodecane, ($\Phi_{oil}$= 0.5), at different pH-values: Region 1 (pH 1.76 and pH 3.74), where solutions are clear; Region 2, during coacervate formation (pH 5.09 and pH 5.72) and in Region 3 (pH 6.35), on the plateau of coacervation. For all samples, manual shaking is insufficient to achieve efficient homogenization, whereas high-speed homogenization gives interesting results, as shown in Figure 1. Following the high-speed homogenization, samples *off-plateau* in Region 1 separate into two distinguished phases (oil and aqueous phase). Samples in Region 2 and 3 are biphasic: a turbid aqueous phase on the bottom and, on top, a white and dense emulsion, which after about four hours at rest, holds its weight upon vial inversion. By adding water or oil, one can easily determine the emulsion's type (o/w or w/o). In this case, water dilutes the emulsion phase, whereas dodecane only increases the volume of the oil layer on the top of the sample, suggesting that the present emulsions type is o/w, in a good agreement with the Finkle rule, which predicts that the continuous phase of the preferred emulsion is the one in which the stabilizer is preferentially dispersed.[52] Reference emulsions of o/SL and o/CHL ($\Phi_{oil}$= 0.5, Figure 1) prepared at pH > 5 form a fluid emulsion, which flow upon vial inversion and destabilizes over time. These observations show that the pH values of SL-PEC, therefore the coacervation stage, play a crucial role in emulsification process. This hypothesis is based on the assumption that neither the coacervates structure nor their composition is significantly affected during the emulsification process. This hypothesis will be discussed later on.

The preliminary experiments shown in Figure 1, underline the effect of the mixture's pH, i.e., the coacervation stage, on the emulsion's macroscopic behavior. One can observe that o/w emulsions are formed with a turbid aqueous bottom phase. The residual turbidity, related to the presence of coacervates after emulsification, highlights an excess of coacervates compared to the amount required to stabilize the o/w interfaces.

To study the influence of the oil content on the emulsion type and macroscopic behavior, the oil volume fraction ($\Phi_{oil}$) is varied from 0 to 0.75, and the appearance of the emulsions is shown in Figure 3. In absence of oil (at $\Phi_{oil}$= 0), solution's turbidity originates from the coacervates' presence in the solution. The progressive increase in $\Phi_{oil}$ leads to an increase in the emulsion volume and a decrease in the lower phase volume (coacervate phase). For $\Phi_{oil}$ up to 0.6, creaming is always present, but it disappears at $\Phi_{oil}$= 0.7. For all samples at $\Phi_{oil}$<

0.7, an estimation of the residual volume of the aqueous phase, by lower phase weighing, highlights that $\Phi_w$ is always about 0.3, which is in agreement with the fact that $\Phi_{oil}$= 0.7 is the optimal o/w volume ratio able to emulsify the solution entirely. This observation is in good agreement with a previous study on emulsion stabilized using poly(N-isopropylacrylamide-co-methacrylic acid) microgels, and where authors attributed the emulsion creaming to an excess of microgels in the solution.[53]

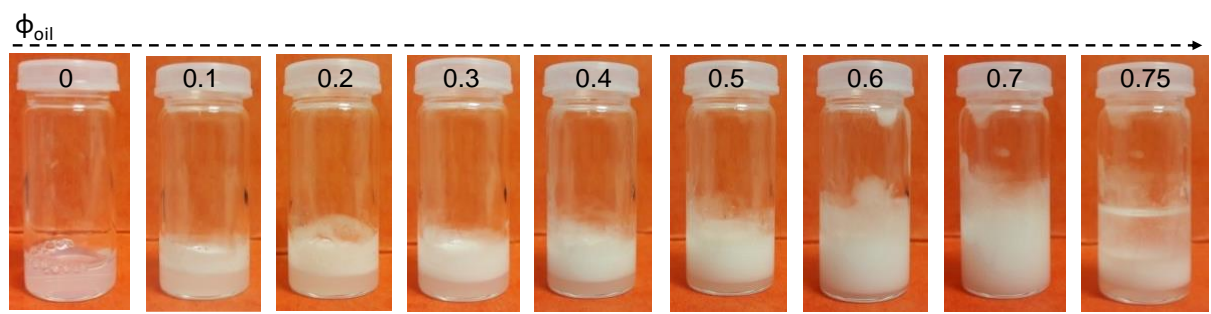

**Figure 3 -** Increasing $\Phi_{oil}$ in an o/w emulsion. The aqueous phase, at pH= 5.74 (strong turbidity plateau in Figure 1), contains a solution of complex coacervates composed of SL and CHL ($C_{SL}$= 5 mg/mL; $C_{CHL}$= 1.4 mg/mL). The pictures are taken four hours after emulsion preparation.

At $\Phi_{oil}$= 0.75> $\Phi_{max}$= 0.74, which is defined as the maximum volume fraction of uniform non-deformable spheres when packed most efficiently,[54] three phases are observed: a large oil phase, a thin emulsion phase, and a thin aqueous phase. However, an optimization of the emulsification process by increasing the homogenization speed and by gradual incorporation of the oil phase during the emulsification process might result in a stable High Internal Phase Emulsion, HIPE ($\Phi_{oil}$ > 0.74). Although HIPE systems' development is of great importance for a wide range of applications,[55,56] the latter is beyond the current study's scope.

Based on these observations, more advanced characterizations are conducted on o/w stabilized by either SL-CHL and SL-PLL complexes and prepared at different pH values and with a selected oil volume fraction ($\Phi_{oil}$) of 0.7.

The imaged droplets and respective size distribution of o/w emulsions, prepared using either SL-CHL and SL-PLL aqueous phase, are shown in Figure 4. SL-CHL system, in Region 2 (pH 5.06, Figure 4a), shows that oil droplets are not aggregated and they can be easily isolated from one another. The size distribution of their size can be fitted by a log-normal function with a mode at about 7 µm. *On-plateau*, at pH 6.35 (Figure 4b), oil droplets exhibit a similar size distribution with a mode at around 7 µm, but a large number of droplets with a diameter < 3 µm form aggregates which are difficult to disperse and to measure accurately using optical microscopy. Thus, small pH variation does not seem to significantly influence the average oil droplet size distribution above the micron (size distribution in Region 2 and Region 3 are similar); however, when coacervates are prepared, *on-plateau*, strong aggregation and small size are common features.

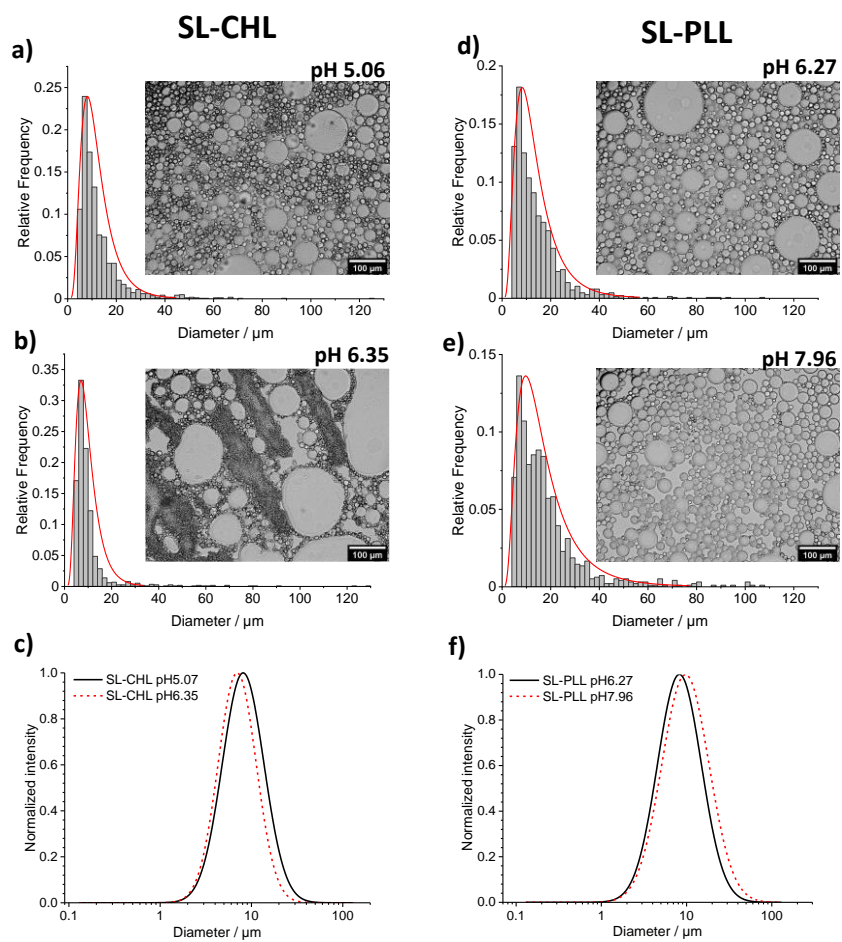

**Figure 4 - Optical microscope images of o/w emulsion ($\Phi_{oil}$= 0.7) stabilized by a solution of complex coacervates composed of (a-b) SL and CHL ($C_{SL}$= 5 mg/mL; $C_{CHL}$= 1.4 mg/mL) at pH= 5.06 (off-plateau, Figure 1), pH= 6.35 (on-plateau, Figure 1) and (d-e) SL and PLL ($C_{SL}$= 5 mg/mL; $C_{PLL}$= 2 mg/mL) at pH= 6.27 and pH= 7.96 (both pH values are chosen to be on the strong turbidity plateau, see Figure S 1). For all samples, the fitted log-normal function, is plotted on c) for SL-CHL and f) for SL-PLL.**

A pH-sensitive and efficient, fully bio-based emulsifier composed of SL-CHL green coacervates is obtained at this stage. It was previously demonstrated that chitosan polysaccharide is considered, in its insoluble form above pH > 6.5, as an efficient Pickering emulsion stabilizer.[47] Moreover, several chitosan-based stabilizers were previously reported. [57]

Therefore, to demonstrate that emulsion stabilization is a common feature of SL-PEC complex coacervates, we tested the SL-PLL system's emulsification properties. This is also motivated by the broader coacervation plateau of SL-PLL, from pH 5.7 to pH 8.5 (Figure S 1), compared to SL-CHL and is related to the difference between the pKa values of SL and those of PLL (~10.5) and CHL (~6.5) and to the insolubility of CHL at pH> 7.[42] Optical microscopy images in Figure 4c,d show a homogeneous distribution of well-dispersed oil droplets stabilized by SL-PLL complex coacervates prepared at the beginning (pH 6.27) and the end (pH 7.96) of the corresponding SL-PLL *plateau*. The mode of the distribution lies between 7 µm and 10 µm and it does not seem that pH has any significant influence neither on size nor on aggregation once the system is *on-plateau*. Comparing both SL-PEC emulsions using optical microscopy, one could observe that oil droplets stabilized by SL-PLL and SL-CHL are of comparable sizes.

### 3.2. Rheological properties of the o/w emulsions

All o/w emulsions prepared *on-plateau* region form a three-dimensional network that does not flow upon the vial inversion test. To quantify the mechanical properties of the obtained emulsions, oscillatory shear rheological characterization is performed. The shear strain and frequency dependence of the storage ($G'$) and loss ($G''$) moduli are shown in Figure 5. The shear strain dependence of $G'$ and $G''$ of o/w emulsion stabilized with SL-CHL and SL-PLL and prepared at different pH values are shown in Figure 5a and Figure 5c, respectively. Both systems displayed the same typical strain-softening behavior with constant values of $G'$ and $G''$ as shear strain increase until a critical shear strain ($\gamma_c$), indicating the limit of the linear viscoelastic regime (LVER) and from which both moduli decrease abruptly.

It's interesting to note that the LVER range of o/w stabilized by SL-CHL is pH-dependent and, more precisely, on the coacervation region. At pH= 5.06, in Region 2, the linear domain extends to $\boldsymbol{\gamma_c}$ ~ 0.83 % and $\boldsymbol{\gamma_c}$ ~ 30 % at pH 5.70 and pH 6.35, *on plateau* (Figure 5). $\boldsymbol{\gamma_c}$ increases with pH, and it shows that emulsion gels are more sensitive to shear at *off-plateau*. While the LVER of o/w stabilized by SL-PLL (Figure 5c) extend to $\boldsymbol{\gamma_c}$ ~ 3 % and $\boldsymbol{\gamma_c}$ ~ 4 % at pH 6.27 and pH 7.96, respectively.
The frequency sweep experiments (Figure 5b,d) show that $G'(\omega)$ is higher than $G''(\omega)$ all over the angular frequency range confirming that the emulsions have predominantly elastic rather than viscous character. The $G'(\omega)$ spans between 113 - 135 Pa and between 48-56 Pa for o/w emulsions stabilized with SL-CHL and SL-PLL; respectively. The rheological measurements of the concentrated emulsion ($\Phi_{oil}$= 0.7) could be assumed to an indirect characterization of the o/w interface strength. In fact, for highly concentrated emulsion ($\Phi_{oil} > \Phi^* = 0.64$, where $\Phi^*$ is the random close packing), the elasticity of the emulsion is governed by the energy storage at the interfaces.[58]

The frequency sweep experiments highlight nearly frequency-independent storage moduli ($G'(\omega) \propto \omega^\alpha$ with $\alpha < 0.08$) with $G'$ varying between 40 - 200 Pa for a series of rheological experiments conducted on o/w emulsions for $\Phi_{oil}$ = 0.7. The variation of $G'$ values from one sample to another could be related to a number of concomitant factors; theoretical models show that the storage modulus ($G'$) of concentrated emulsions is inversely proportional to the emulsion droplet radius ($r$) but directly proportional to the interfacial tension ($\sigma$) and $\Phi(\Phi-\Phi^*)$ (Princen:[59] $G' \propto 1.77\Phi^{1/3}(\Phi - \Phi^*)\sigma/r$ and Mason *et al.*:[58] ($G' \propto \Phi(\Phi - \Phi^*)\sigma/r$). In the present work, SL-CHL and SL-PLL exhibit similar air-water interfacial surface tension of 32.7 and 32.6 mN/m, respectively (Figure 2). Therefore, one possible key parameter could then be the emulsion droplet size and its broad size distribution. As shown in Figure 4, the emulsions are mainly composed of droplets between 10 and 20 µm, coexisting with large droplets as well as droplets <3 µm. These could be embedded in the empty space between the larger droplets leading to an increase of the final viscoelastic properties.[60] Preliminary tests, not shown here, confirm an effect of the droplet size on the elastic properties of o/w emulsion gels: reducing the droplet size to less than a micron provided $G'$ in the order of 1 kPa.

Table 2 shows the comparative elastic properties of both classical (surfactant-based) and Pickering (particle-based) emulsions. For comparable volume fraction, the values of $G'$ are in the same order of magnitude of concentrated emulsions stabilized by surfactant adsorption [61] or by forming a continuous protein protective layer.[62] Concentrated emulsions stabilized by a Pickering effect generally demonstrate, for a similar volume fraction, a higher $G'$ by at least one order of magnitude than classical concentrated emulsion.[63] The particles' origin and role in increasing the interface elasticity are system-dependent and not predictable and could depend on several factors related to the properties of the used particles, e.g., size, shape, rigidity, and deformability, and the interaction between them. However, concentrated emulsion stabilized by a Pickering effect using solid particles like hydrophobic starch,[64] kafirin,[65] and cellulose nanocrystals or nanofibrils[66] or using soft and deformable particles like microgels demonstrates comparable gel-like properties[67] with $G'$ being in the order of 100 Pa at 1 Hz. The emulsions prepared in this study by using SL-CHL and SL-PLL complex coacervates as stabilizers also show a similar gel-like behavior with comparable $G'$.

**Table 2 – Comparison between the typical storage moduli of emulsions measured using small amplitude oscillatory shear rheology at a frequency of 1 Hz.**

| Stabilizer | Oil fraction | $G'$ / Pa | Note | Stability | Ref |
|---|---|---|---|---|---|
| surfactant | 0.7 | < 1 | | Yes | [61] |
| fish gelatin | 0.7 | 2-20 | According to gelatin concentration | Yes | [62] |
| β-lactoglobulin B | 0.6 | 1-200 | According to physicochemical conditions | Yes | [68] |
| starch | 0.67 | 70 | | Yes | [64] |
| kafirin (protein) | 0.7 | 100 | | Yes | [65] |
| cellulose nanocrystals (CNC) | 0.01 (1 wt%) | 100 | > 0.3 wt% CNC | Yes | [66] |
| Microgels | 0.7 | 100 | | Yes | [67] |
| *Sophorolipid-based PESC coacervates* | *0.7* | *100* | | *pH-dependent* | *This work* |

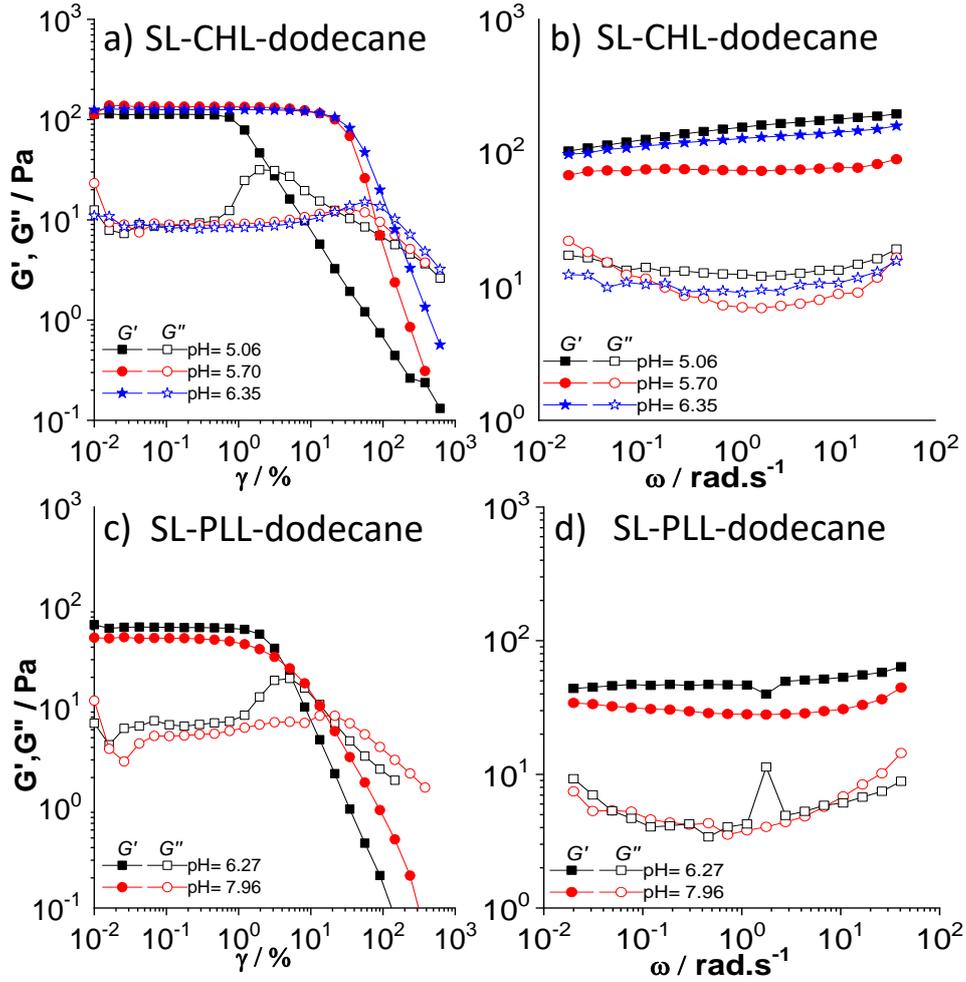

**Figure 5** - Storage (G′) and loss (G″) moduli as a function of (a,c) strain, $\gamma$, and (b,d) oscillation frequency, $\omega$, for (a,b) SL-CHL and (c,d) SL-PLL o/w emulsion ($\Phi_{oil}$= 0.7). Experiments show the dependency on the pH of the aqueous phase. $C_{SL}$= 5 mg/mL; $C_{PLL}$= 2 mg/mL; $C_{CHL}$= 1.4 mg/mL. Strain sweep measurements are carried out at 6.28 rad/s while frequency sweep is done at $\gamma$= 0.3 %. In a,b), pH 5.06 is *off-plateau* (Figure 1), pH= 5.70, 6.35 are *on-plateau* (Figure 1). In c,d) both pH are *on-plateau* (Figure S 1).

### 3.3. pH-responsive and mechanical recovery properties

Both SL-CHL and SL-PLL-based o/w emulsions are stable over months at room temperature at complete rest or even after manual shaking, mimicking an uncontrolled mechanical solicitation event, like transport. Based on this observation, step-strain cycles experiments are performed on an o/w emulsion stabilized by SL-PLL complex *on-plateau* (Figure 6) to quantify the mechanical stability and potential recovery time of the emulsions. These experiments consist of applying a small shear strain within the LVER ($\gamma$ = 0.3%) to the sample for the first 30 minutes to evaluate its initial rheological properties ($G'_i$= 53 Pa). The emulsion is destructured for 2 minutes by applying a large shear strain ($\gamma$ = 20%) out of the LVER (see Figure 5c), when $G' <$ 10 Pa and $G' < G''$, characteristic of liquid behavior. The sample is allowed to recover its properties for 30 min under a strain set within the LVER ($\gamma$ = 0.3%). The first measurement after recovery indicates that the emulsion has recovered its gel behavior ($G' > G''$) and 60% of its initial $G'$. After 30 min, the gel has recovered 75% of its initial storage modulus. Five cycles of large and small strains (Figure 6) show that the emulsion has the same behavior and it recovers the same percentage of $G'$ after each cycle. After the last cycle, recovery is allowed for one hour and emulsion recovers almost 85% of its initial $G'$, as well as the same response to strain and frequency sweep (Figure S 2). This experiment demonstrates that the homogenization of a coacervate solution with an oil phase produces a highly stable gel with reversible gelling upon applying an external mechanical strain. This and their ability to restructure quickly after destructuring are evidence of the stretching of the oil-water interfaces.

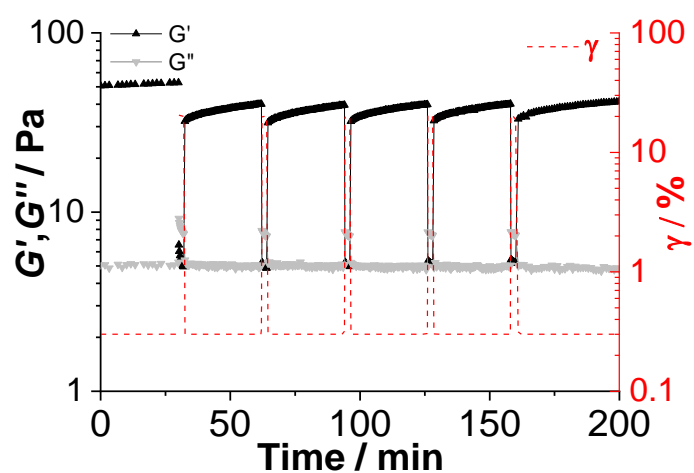

**Figure 6** - Evolution of storage ($G'$) and loss ($G''$) moduli during step-strain experiments on a oil-in-water emulsion *gel* ($\Phi_{oil}$= 0.7). The aqueous phase (pH= 6.27, *on-plateau*, Figure S 1) contains a solution of SL-PLL complex coacervates ($C_{SL}$= 5 mg/mL; $C_{PLL}$= 2 mg/mL).

Based on the pH-reversibility of complex coacervation formation, emulsions' mechanical properties are also tested against *in-situ* pH variation. Figure 7 shows the mechanical properties and optical microscopy of o/w emulsions stabilized by SL-CHL and SL-PLL complexes and characterized by an initial $G'$ between 80 Pa and 100 Pa, respectively. When the pH is lowered to the *off-plateau* region, a turbid aqueous and an oil phase form after few minutes at rest (image at pH= 1.8 in Figure 7a). Upon increasing pH to its initial *on-plateau* value followed by standard homogenization, it is possible to recover an emulsion gel with comparable mechanical properties. This behavior is observed for both PEC-based o/w emulsions, and it can be visioned in Video 1 as multimedia support in the Supplementary

Material. This observation points out the crucial role of complex coacervates in the stabilization mechanism of the emulsions. Another argument in favor of this hypothesis is that the emulsions stability depends on parameters that strongly affect the complex coacervates; that is the state of emulsions and coacervates seems to be closely connected. The optical microscope images and the corresponding droplet size distribution of o/w stabilized by SL-CHL coacervates at pH 5.7, after preparation and following destabilization - re-emulsification by pH change treatment, are shown in Figure 7b-c. Optical microscope images of the initial o/w emulsion highlight the main droplet population with a diameter below 3 µm and a second minor population of larger droplets, between 15 and 35 µm. While the obtained emulsion after pH change treatment shows two populations with a diameter below 3 µm and a diameter between 10 and 20 µm. Fitting the droplet size distribution to a log-normal distribution function highlights a similar mode around 3 µm for both emulsions. The variation of the emulsions droplet size distribution, before and after pH change treatment, could explain the small variation of the resulting rheological properties and suggest potential rearrangement of the coacervates phase at the o/w interface.

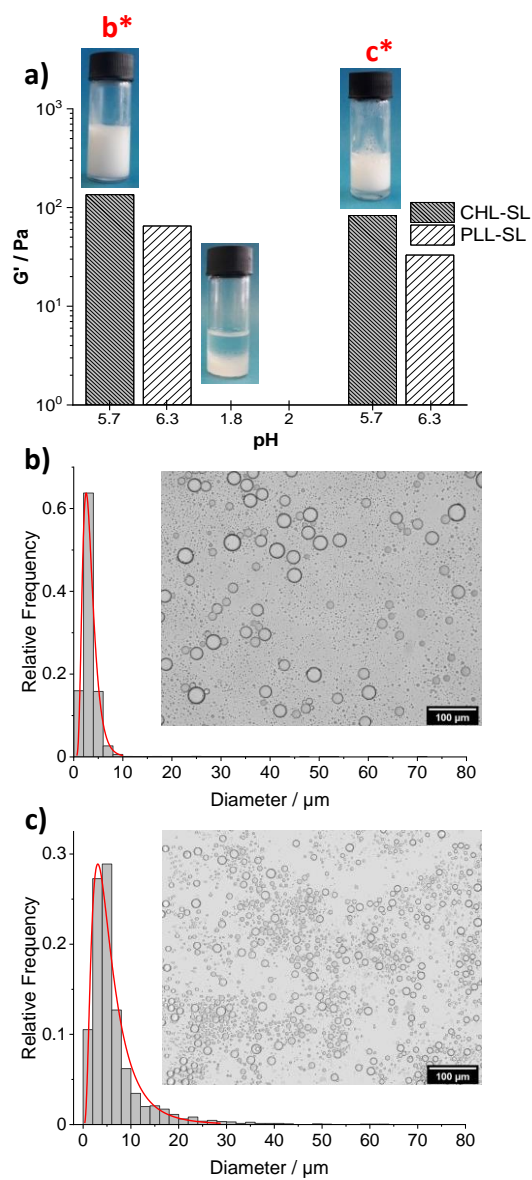

**Figure 7** - Comparison of the storage modulus ($\omega$= 6.28 rad/s, $\gamma$= 0.3 %) of o/w emulsion gels ($\Phi_{oil}$= 0.7) containing complex coacervate solutions composed of a) SL-CHL ($C_{SL}$= 5 mg/mL; $C_{CHL}$= 1.4 mg/mL, pH= 5.70) and b) SL-PLL ($C_{SL}$= 5 mg/mL; $C_{PLL}$= 2 mg/mL, pH= 6.27) after a pH-change treatment. Both emulsion gels are prepared *on-plateau* (Figure 1 for SL-CHL and Figure S 1 for SL-PLL) at the optimum pH; pH is lowered below the complex coacervation region: phase separation occurs; pH is increased back to its optimum value on the plateau, emulsions are homogenized again according to the procedure described in the materials and methods section: emulsion gels are obtained again. Symbols b* and c* indicate the samples used for optical microscopy images and emulsion droplets distribution of the o/w stabilized by SL-CHL complex, respectively at pH 5.7 b) after emulsion preparation and c) after pH change treatment. A full video of the pH-dependent emulsification process is given in Video 1 as a Multimedia support in the Supporting Information.

## 3.4. Mechanism of stabilization of emulsion gels: Pickering effect

Emulsions stabilized by SL-CHL and SL-PLL form stable, pH-reversible solid-like systems, compared to those stabilized by either SL or CHL (Figure 1) prepared as references and which behave as liquid fluids. We have also shown the crucial impact of pH, which has to be set *on-plateau* of the SL-PEC turbidity curve, where SL and PEC form complex coacervates, which play an essential role in the stabilization of the oil-water interface. Complex coacervates have a spheroidal morphology with size ranging from about 50 nm to several hundred nanometers, depending on the composition and pH.[42] The remaining question is the mechanism underlying emulsion's stability coacervates. The latter have an amphiphilic character and are surface-active, as highlighted by the surface tension measurements (Figure 2).

To elucidate the emulsion stabilization mechanism by the SL-PEC complex coacervates, it's essential to discuss the behavior of interacting surfactants-PEC systems at the interface. The complexation of surfactant and polymers and their adsorption behavior on water /air or solid surface were critically reviewed [69].

The behavior of oppositely charged polymer-surfactant at air-water interface was previously investigated using surface tension measurements, neutron reflectivity or ellipsometry [70–72]. It showed that poly(sodium-4-styrene sulfonate) (PSSNa) - dodecyltrimethylammonium bromide (DTAB) and poly(2-acrylamido-2-methyl-1-propanesulfonic acid) (PAMPS)-DTAB or PAMPS- hexadecyltrimethylammonium bromide (CTAB), result in the formation of a multilayer structure at the air/water interface. It's presumed that the interface is composed of an initially adsorbed surfactant monolayer associated with the surfactant-polyelectrolyte complex. The latter are of an identical or a different structure from the complex, which are in bulk. It's assumed that these structures are surfactant micellar-like aggregates adsorbed on the polymer chains, while other studies suggested the formation of interfacial gels or microgels [73]. Chitosan cannot stabilize emulsions when it is water-soluble at acidic pH. To solve this problem, a recent work [48] has shown the beneficial use of SDS with chitosan at acidic pH. Chitosan-SDS stabilize emulsions by forming either insoluble flocs or polymeric surfactant, depending on the molecular weight of the cationic polysaccharide, which preferentially adsorb at the water-oil interface.

Generally, the presence of oppositely charged surfactant - PEC molecules result in their cooperative adsorption at fluid interfaces[74]. However, this synergistic effect tends to disappear when the free surfactant concentration is high enough to compete for the interface's available area [75].

In the present work, pH-induced complex coacervates are formed between a cationic PEC (CHL or PLL) and a bolaform surfactant above its CMC. In this regard, it was previously shown that the air/water interface stabilization for sufficiently high surfactant concentrations is two diffusion-controlled adsorption processes with the first adsorption of free surfactant molecules followed by the adsorption of polymer–surfactant complexes [76]. However, this diffusive control is closely related to the bulk's complexation state as demonstrated for polyethylenimine- sodium dodecylsulfate system and where the interfacial adsorption process (diffusion or non-diffusion controlled) is pH-dependent [77]. Therefore, the sequential adsorption of the free surfactant and surfactant-PEC complexes is not a general feature.

At first glance, it is reasonable to think that the adsorption of SL-PEC complexes will be slower than the single SL and PEC components due to their larger size. However, one should note that during the coacervation process, complex coacervates are formed and are in a thermodynamic equilibrium with diluted phase composed of SL and PEC. Predicting whether the remaining free SL concentration is below or above the CMC is difficult.

Based on the general surfactant-PEC interfacial behavior, two mechanisms seem to be conceivable: i) the complex coacervates reorganize into a continuous protecting layer surrounding the oil droplets,[28] classically described in the literature, or ii) the coacervates adsorb on the oil droplet surface and stabilize the emulsion by a Pickering effect, as shown for emulsions stabilized using polysaccharide-protein like zein-based colloids [22] and chitosan-gliadin coacervates [19]. In the latter scenario, a stabilization by solely the coacervates or a diffusion-controlled adsorption of first SL molecules forming a surfactant monolayer followed by the coacervates structures are both conceivable.

To elucidate the emulsions' stabilization mechanism by SL-PEC complex coacervates, we engaged a microscopy study using CLSM and cryo-SEM. These techniques are selected because they allow exploration of the as-prepared o/w interface without sample drying, thus avoiding artifacts.

Figure 8 presents a series of CLSM images of o/w emulsion stabilized by SL-PLL complexes and labeled with FITC-PLL (green, single labeling, Figure 8a-c) or both FITC-PLL and Rhodamine-labeled lipids (green/red, double labeling, Figure 8d-f). The single labeling CLSM image (Figure 8a) shows that green fluorescence (FITC-PLL and more likely PLL-SL coacervates) is visible only at the oil−water interface with the formation of an interconnected network due to the bridging of the emulsion droplets. Figure 8b shows the top surface of an oil droplet covered with discrete green-fluorescent particles, the size of which is contained between 0.5 µm and 1.5 µm (Figure 8c) and with an inter-particles distance (center-to-center) between 2.0 µm and 3.4 µm, as determined by the denoised magnification and corresponding Fast Fourier Transform in Figure 8c (magnification and corresponding FT of zones (2) and (3) of Figure 8a are shown in Figure S 3). The largest size of SL-PLL complex coacervates in pure water *on-plateau* is estimated to be between 0.5 µm and 1 µm, as observed before by optical microscopy and dynamic light scattering,[42] in good agreement with the particle size on the oil surface. Besides, the green fluorescence, indicative of FITC-labeled PLL localization, confirms that the particles contain the polycation. Possible confusion between adsorbed oil droplets and complex coacervates is minimized by the strong difference in contrast between obviously adsorbed oil droplets (core-shell structure, black arrows in Figure S 4-4b) and coacervates (dense green particles, white arrows, in Figure S 4-4a). To verify that the particles also contain a micellar component, we have simultaneously labeled SL micelles using a 1:500 molar ratio of a rhodamine-labeled monounsaturated phospholipid (Rho-PE) against SL, with the hypothesis that such a substantial dilution has no impact neither the coacervation process nor in the emulsion stabilization. This hypothesis is validated because one can obtain an o/w emulsion containing FITC (green) and rhodamine (red) labeling simultaneously, respectively identifying the PEC and micellar phases.

Figure 8d-f, corroborated by additional experiments shown in Figure S 5 and presented in Video 2 and Video 3 as a Multimedia support in the Supporting Information, show droplets, of which the surface contains both red- (Figure 8d, Figure S 5a, Figure S 5d) and green- (Figure 8e, Figure S 5b, Figure S 5f) labeled colloids. Their specific co-localization, shown by the white/yellowish particles and interfaces in Figure 8f, Figure S 5c, Figure S 5g, proves that the colloids are composed of PLL and SL micelles. Interestingly, the continuous green background in Figure 8a,e and the homogeneous red surface of the droplets respectively indicate that the aqueous medium contains an excess of PLL, while the oil droplet's surface is rich in SL micelles. Two events could possibly explain this evidence: 1) the actual lipid-to-polyelectrolyte ratio may not be optimized, meaning that unbound polyelectrolyte and lipid will have different affinities, respectively, for water and oil-water interface; 2) the lipid-to-polyelectrolyte ratio is indeed optimized, but a fraction of the PEC and lipids dissociates in the oil-water medium. Further experiments are required to draw clear-cut conclusions. Additional insight into the water-oil interface is provided by complementary cryo-SEM experiments, of which the resolution is much higher than CLSM, limited to around 250-300 nm.

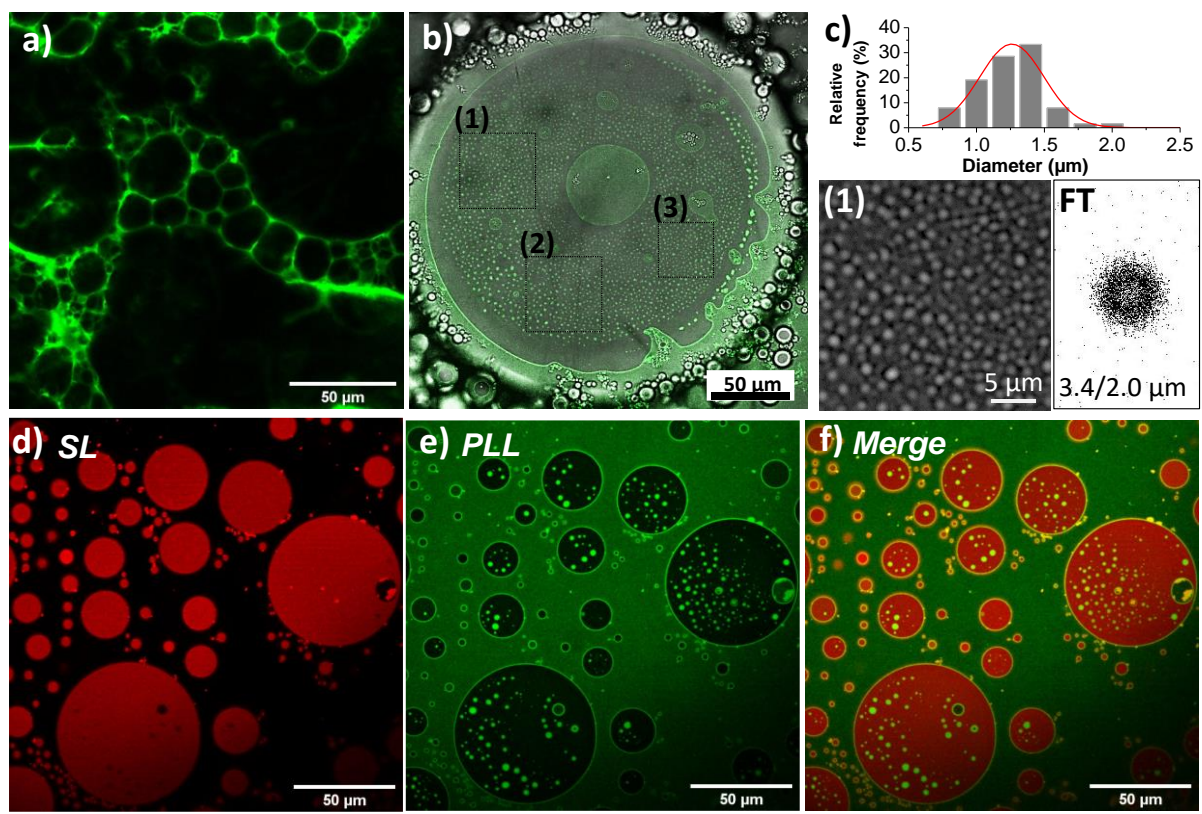

**Figure 8** – CLSM images from single labeling experiments: a) o/w emulsion at pH= 6.27 b) merged white and PLL-FITC green channel of a typical gelled emulsion composed of water and SL-PLL complex coacervates prepared at a plateau pH value (Figure 1, Figure S 1). Green particles correspond to complex coacervates containing FITC-labeled PLL. c) Size distribution of the complex coacervates, magnification of region (1) and corresponding Fourier Transform observed in a) (regions (2) and (3) are shown in the Supporting Information). d-f) CLSM images from double labeling experiments: d) the Rho-PE is given in red, e) the PLL-FITC in green and f) merged (red and green) channel CLSM images of a typical o/w emulsion stabilized by SL-PLL prepared at pH 5.64. For all experiments, the selected z-plane corresponds to the top surface of the oil droplets. Details on each experiment's methodology and composition are given in Table 1 in the materials and methods section. A full set of CLSM experiments is provided in Video 2 and Video 3 as a Multimedia support in the Supporting Information.

High-resolution cryo-SEM experiments are shown in Figure 9 and Figure S 6; Figure 9a-b (o/w emulsion stabilized by SL-PLL) shows the negative imprinting of a droplet, accidentally removed during sample preparation. The red line delimits the bulk water solution (right-hand-side) from the cavity that contained the droplet (left-hand side). One can observe a continuum of densely aggregated colloidal particles from the bulk and surrounding the droplet's imprint. Similar observations are also provided in Figure S 6a,b, imaged on another sample sector. Such observation is in good agreement with what was observed for Pickering water-in-oil emulsion stabilized by poly(N-isopropylacrylamide) microgels.[78] On the contrary, Figure 9c-d show a single oil droplet massively surrounded by discrete colloidal particles. Figure 9d shows, in particular, a closer look at a particle surface covered with colloids. A similar phenomenon is also observed in another sector of the same sample (Figure S 6c,d). The estimated size of these colloids is in the order of 50 nm, in good agreement with the smallest complex coacervates observed before by cryogenic transmission electron microscopy. [42] Cryo-SEM was previously performed on various Pickering emulsion systems. A similar oil droplet surface covered with particles prepared from surfactants[79], proteins[65], and soft deformable microgels[80] have been observed.

As expected, cryo-SEM provides a complementary view of the droplet surface, showing that it is composed of small colloids (<< 1 μm) undetectable by CLSM. Based on CLSM and cryo-SEM observations, we observe that the o/w interface is characterized by a massive presence of spheroidal colloids of typical complex coacervates size. For this reason, we can exclude the stabilization mechanism through the rearrangement of the coacervates during the emulsification process into a continuous layer surrounding the oil droplets.[81] Although the elucidation of emulsion stabilization mechanism through PEC complex remains challenging,[17,82,83] we are prone here for a colloidal stabilization, thus identifying a typical Pickering effect. This assumption is also coherent with the gel properties ($G' > G''$) of the emulsions, where $G'$ is in the order of 100 Pa, as typically found for Pickering emulsions at comparable oil fractions (Table 2).

Traditionally, the Pickering stabilization mechanism is generally associated with the presence of rigid particles at the oil-water interface and which generally share some common feature like the partial particles wettability by both the water and oil phase [84]. Recent work has shown a Pickering effect even when soft (Young's modulus in the order of kPa) colloids are employed like microgels [9,12]. In contrast with solid particles (Young's modulus in the order of GPa), microgels are fully swollen by the water phase but still adsorb at the oil droplet surface. Moreover, an additional factor, particle deformability, plays an essential role in the emulsion stabilization mechanism. Other soft colloid systems like PEC-PEC and polysaccharide-protein complexes coacervates seem to share some standard features with the microgels systems [17,19,23]. The stabilization of the o/w interface in the present work then joins the latter category, as by definition a complex coacervate is a liquid-liquid phase separation, hence with an expected Young's modulus less than the kPa.

All in all, PESC coacervates composed of a mild surfactant with tunable charge can be used as Pickering stabilizers for o/w emulsions without major changes in their colloidal structure, as long as the emulsion is prepared in the region of complex coacervation. If one cannot exclude partial diffusion of the surfactant and the polyelectrolyte respectively towards the o/w interface and water phase, the overall morphology and size of the complex coacervates is comparable to what it is found in solution. The elastic properties of the emulsions are then comparable to those Pickering emulsions obtained by both hard and soft colloids.

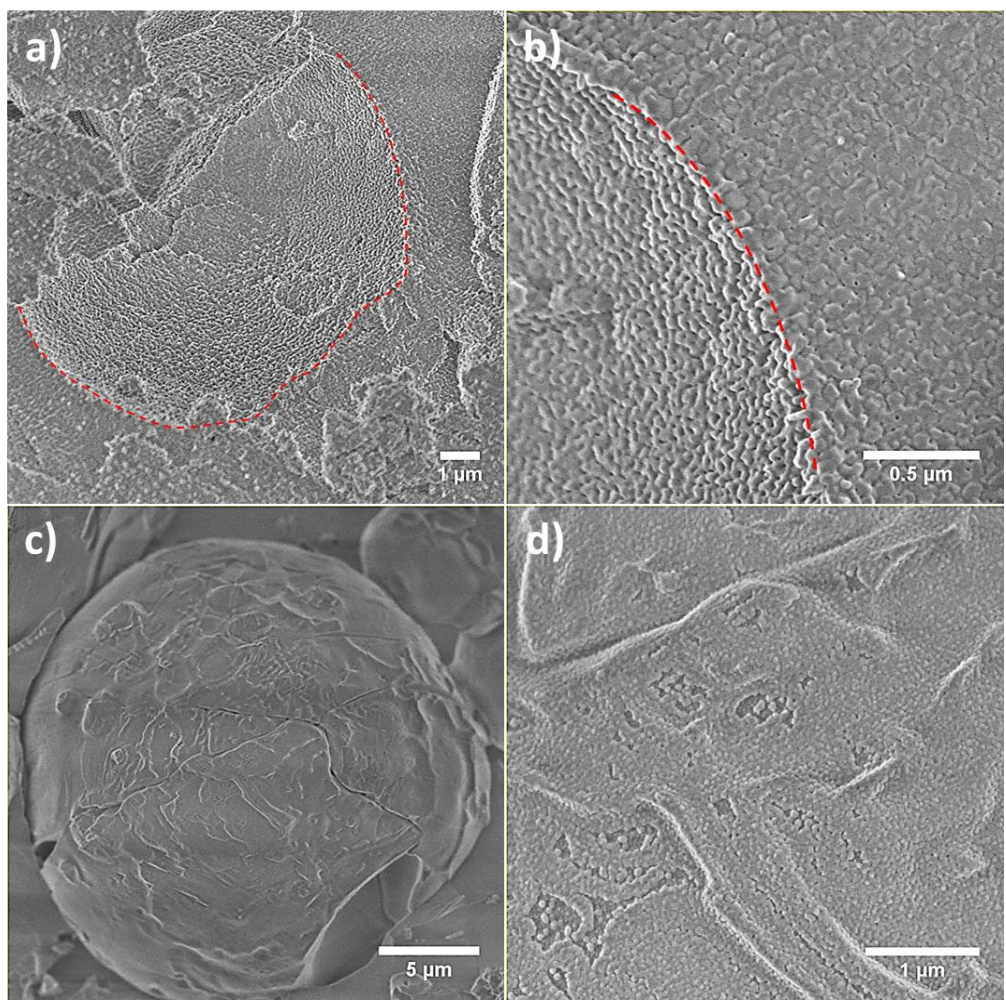

**Figure 9** – Cryo-SEM images of o/w emulsions ($\Phi_{oil}$= 0.7) and an aqueous solution of complex coacervates composed of (a-b) SL and PLL ($C_{SL}$= 5 mg/mL; $C_{PLL}$= 2 mg/mL) at pH= 6.55 (on-plateau, Figure 1), (c-d) SL and CHL ($C_{SL}$= 5 mg/mL; $C_{CHL}$= 1.4 mg/mL) at pH= 6.22 (on-plateau, Figure S 1).

## 4. Conclusion

In this work, we show that polyelectrolyte-surfactant complex (PESC) coacervates act as efficient emulsion stabilizers through a Pickering effect, thus adding PESCs coacervates to the family of soft colloidal Pickering emulsion stabilizers along with microgels, PEC-PEC or PEC-protein complexes. PESC coacervates do not disassemble at complex o/w interfaces and they actually show an amphiphile character, necessary to settle at the o/w interface. The use of sustainable sophorolipid biosurfactants demonstrate that micelles with a variable charge and less marked amphiphilic character [31] can be successfully associated to natural polyelectrolytes. Pickering emulsions based on this new class of PESC coacervates are entirely biobased and ecofriendly but also stable and pH-responsive, two characteristics that could be reached with classical petrochemical surfactants.

o/w emulsions are prepared by varying the oil volume fraction and are stable up to $\Phi_{oil} = 0.7$, larger, than the oil fraction stabilized using polysaccharide-surfactant complexes, like cellulose – surfactant (50%) [85–87], chitosan-sodium dodecylsulfate [48] and chitosan-sodium sodium decane-1-sulfonate surfactant (~ 26 %) [49]. The system's typical rheological signature corresponds to a gel with $G'$ in the order of 100 Pa, typical of Pickering emulsions obtained with proteins, [65,68] microgels [67] or cellulose nanocrystals [66] (Table 2). The Pickering effect is shown by a combination of CLSM and cryo-SEM, highlighting the massive presence of coacervates at the oil/water interface.

The pH-responsive feature of the SL-PEC stabilizer and their biobased nature make it an attractive candidate for a wide range of applications and where temporary controlled stability is required, like the on-demand release of functional bioactive molecules for biological applications. However, the free surfactant and PEC molecules' implication level in the stabilization process and understanding the adsorption mechanism of the complex coacervates at the oil surface droplets and their potential structural modification, like deformation remains a significant challenge and should stimulate future investigation.

## 5. Acknowledgments

Research leading to these results received funding from the European Community's Seventh Framework Programme (FP7/2007-2013) under grant agreement no. Biosurfing/ 289219. Estelle Anceaume and Julien Dumont (Center of interdisciplinary research in biology imaging facility, Collège de France, Paris, France) are kindly acknowledged for confocal image acquisition. Cécile Monteux (CNRS, Laboratory Soft Matter Science and Engineering, ESPCI, Paris, France) is kindly acknowledged for helpful discussion.

## 6. References

[1]   S. Friberg, K. Larsson, J. Sjoblom, Food emulsions, CRC Press/Taylor & Francis, 2003. https://books.google.fr/books?hl=de&lr=&id=2XvHZtRRA7sC&oi=fnd&pg=PP1&dq=emulsions+for+food+applications&ots=KbltwtP3sW&sig=ph-kfsgJo3bmDaOcNtWnKGOIfwc (accessed January 10, 2020).

[2] P. Boonme, Applications of microemulsions in cosmetics, J. Cosmet. Dermatol. 6 (2007) 223–228. https://doi.org/10.1111/j.1473-2165.2007.00337.x.

[3] E. Bouyer, G. Mekhloufi, V. Rosilio, J.L. Grossiord, F. Agnely, Proteins, polysaccharides, and their complexes used as stabilizers for emulsions: Alternatives to synthetic surfactants in the pharmaceutical field?, Int. J. Pharm. 436 (2012) 359–378. https://doi.org/10.1016/j.ijpharm.2012.06.052.

[4] W. Ramsden, Separation of solids in the surface-layers of solutions and 'suspensions' (observations on surface-membranes, bubbles, emulsions, and mechanical coagulation).—Preliminary account, Proc. R. Soc. London. 72 (1903) 156–164. https://doi.org/10.1098/rspl.1903.0034.

[5] S.U. Pickering, Emulsions, J. Chem. Soc. Trans. 91 (1907) 2001–2021. https://doi.org/10.1039/CT9079102001.

[6] B.P. Binks, Particles as surfactants - Similarities and differences, Curr. Opin. Colloid Interface Sci. 7 (2002) 21–41. https://doi.org/10.1016/S1359-0294(02)00008-0.

[7] Y. Chevalier, M.A. Bolzinger, Emulsions stabilized with solid nanoparticles: Pickering emulsions, Colloids Surfaces A Physicochem. Eng. Asp. 439 (2013) 23–34. https://doi.org/10.1016/j.colsurfa.2013.02.054.

[8] S. Fujii, E.S. Read, B.P. Binks, S.P. Armes, Stimulus-responsive emulsifiers based on nanocomposite microgel particles, Adv. Mater. 17 (2005) 1014–1018. https://doi.org/10.1002/adma.200401641.

[9] W. Richtering, Responsive emulsions stabilized by stimuli-sensitive microgels: Emulsions with special non-pickering properties, Langmuir. 28 (2012) 17218–17229. https://doi.org/10.1021/la302331s.

[10] J. Tang, P.J. Quinlan, K.C. Tam, Stimuli-responsive Pickering emulsions: Recent advances and potential applications, Soft Matter. 11 (2015) 3512–3529. https://doi.org/https://doi.org/10.1039/C5SM00247H.

[11] J. Masliyah, Z.J. Zhou, Z. Xu, J. Czarnecki, H. Hamza, Understanding Water-Based Bitumen Extraction from Athabasca Oil Sands, Can. J. Chem. Eng. 82 (2004) 628–654. https://doi.org/10.1002/cjce.5450820403.

[12] S. Wiese, A.C. Spiess, W. Richtering, Microgel-stabilized smart emulsions for biocatalysis, Angew. Chemie - Int. Ed. 52 (2013) 576–579. https://doi.org/10.1002/anie.201206931.

[13] Z. Zhang, M. Cheng, M.S. Gabriel, Â.A. Teixeira Neto, J. da Silva Bernardes, R. Berry, K.C. Tam, Polymeric hollow microcapsules (PHM) via cellulose nanocrystal stabilized Pickering emulsion polymerization, J. Colloid Interface Sci. 555 (2019) 489–497. https://doi.org/10.1016/j.jcis.2019.07.107.

[14] E. Kizilay, A.B. Kayitmazer, P.L. Dubin, Complexation and coacervation of polyelectrolytes with oppositely charged colloids, Adv. Colloid Interface Sci. 167 (2011) 24–37. https://doi.org/10.1016/j.cis.2011.06.006.

[15] C. Schmitt, S.L. Turgeon, Protein/polysaccharide complexes and coacervates in food systems, Adv. Colloid Interface Sci. 167 (2011) 63–70. https://doi.org/10.1016/j.cis.2010.10.001.

[16] H. Monteillet, F. Hagemans, J. Sprakel, Charge-driven co-assembly of polyelectrolytes across oil–water interfaces, Soft Matter. 9 (2013) 11270–11275. https://doi.org/10.1039/c3sm52241e.

[17] A.M. Bago Rodriguez, B.P. Binks, T. Sekine, Emulsion stabilisation by complexes of oppositely charged synthetic polyelectrolytes, Soft Matter. 14 (2018) 239–254. https://doi.org/10.1039/c7sm01845b.

[18] X. Li, R. de Vries, Interfacial stabilization using complexes of plant proteins and polysaccharides, Curr. Opin. Food Sci. 21 (2018) 51–56.

[19] M.F. Li, Z.Y. He, G.Y. Li, Q.Z. Zeng, D.X. Su, J.L. Zhang, Q. Wang, Y. Yuan, S. He, The formation and characterization of antioxidant pickering emulsions: Effect of the interactions between gliadin and chitosan, Food Hydrocoll. 90 (2019) 482–489. https://doi.org/10.1016/j.foodhyd.2018.12.052.

[20] C.P. Zhu, H.H. Zhang, G.Q. Huang, J.X. Xiao, Whey protein isolate—low methoxyl pectin coacervates as a high internal phase Pickering emulsion stabilizer, J. Dispers. Sci. Technol. 0 (2020) 1–12. https://doi.org/10.1080/01932691.2020.1724801.

[21] Z.K. Zhang, J.X. Xiao, G.Q. Huang, Pickering emulsions stabilized by ovalbumin-sodium alginate coacervates, Colloids Surfaces A Physicochem. Eng. Asp. 595 (2020) 124712. https://doi.org/10.1016/j.colsurfa.2020.124712.

[22] L. Dai, S. Yang, Y. Wei, C. Sun, D.J. McClements, L. Mao, Y. Gao, Development of stable high internal phase emulsions by pickering stabilization: Utilization of zein-propylene glycol alginate-rhamnolipid complex particles as colloidal emulsifiers, Food Chem. 275 (2019) 246–254. https://doi.org/10.1016/j.foodchem.2018.09.122.

[23] W. Xiong, Q. Deng, J. Li, B. Li, Q. Zhong, Ovalbumin-carboxymethylcellulose complex coacervates stabilized high internal phase emulsions: Comparison of the effects of pH and polysaccharide charge density, Food Hydrocoll. 98 (2020). https://doi.org/10.1016/j.foodhyd.2019.105282.

[24] D.S. Williams, S. Koga, C.R.C. Hak, A. Majrekar, A.J. Patil, A.W. Perriman, S. Mann, Polymer/nucleotide droplets as bio-inspired functional micro-compartments, Soft Matter. 8 (2012) 6004–6014. https://doi.org/10.1039/c2sm25184a.

[25] B. Kronberg, K. Holmberg, B. Lindman, Surface Chemistry of Surfactants and Polymers, John Wiley & Sons, Inc., 2014. https://doi.org/10.1002/9781118695968.

[26] M. Gradzielski, I. Hoffmann, Polyelectrolyte-surfactant complexes (PESCs) composed of oppositely charged components, Curr. Opin. Colloid Interface Sci. 35 (2018) 124–141. https://doi.org/10.1016/j.cocis.2018.01.017.

[27] W. Zhao, Y. Wang, Coacervation with surfactants: From single-chain surfactants to gemini surfactants, Adv. Colloid Interface Sci. (2016) 199–212. https://doi.org/10.1016/j.cis.2016.04.005.

[28] B. Lindman, F. Antunes, S. Aidarova, M. Miguel, T. Nylander, Polyelectrolyte-surfactant association—from fundamentals to applications, Colloid J. 76 (2014) 585–594. https://doi.org/10.1134/S1061933X14050111.

[29] S.S. Cameotra, R.S. Makkar, J. Kaur, S.K. Mehta, Synthesis of biosurfactants and their advantages to microorganisms and mankind, Adv. Exp. Med. Biol. 672 (2010) 261–280. https://doi.org/10.1007/978-1-4419-5979-9_20.

[30] D.J. McClements, L. Bai, C. Chung, Recent Advances in the Utilization of Natural Emulsifiers to Form and Stabilize Emulsions, Annu. Rev. Food Sci. Technol. 8 (2017) 205–236. https://doi.org/10.1146/annurev-food-030216-030154.

[31] S. Manet, A.S. Cuvier, C. Valotteau, G.C. Fadda, J. Perez, E. Karakas, S. Abel, N. Baccile, Structure of Bolaamphiphile Sophorolipid Micelles Characterized with SAXS, SANS, and MD Simulations, J. Phys. Chem. B. 119 (2015) 13113–13133. https://doi.org/10.1021/acs.jpcb.5b05374.

[32] U. Rau, S. Hammen, R. Heckmann, V. Wray, S. Lang, Sophorolipids: A source for novel compounds, Ind. Crops Prod. 13 (2001) 85–92. https://doi.org/10.1016/S0926-6690(00)00055-8.

[33] M. Maingault, Use of Sophorolipids and Cosmetic and Dermatological Compositions - WO/1995/034282A., 1995.

[34] S. Lang, E. Katsiwela, F. Wagner, Antimicrobial Effects of Biosurfactants, Eur. J. Lipid Sci. Technol. 91 (1989) 363–366. https://doi.org/10.1002/lipi.19890910908.

[35] G. Ben Messaoud, P. Le Griel, D. Hermida-Merino, S.L.K.W. Roelants, W. Soetaert, C.V. Stevens, N. Baccile, pH-Controlled Self-Assembled Fibrillar Network Hydrogels: Evidence of Kinetic Control of the Mechanical Properties, Chem. Mater. 31 (2019) 4817–4830. https://doi.org/10.1021/acs.chemmater.9b01230.

[36] I.N.A. Van Bogaert, K. Saerens, C. De Muynck, D. Develter, W. Soetaert, E.J. Vandamme, Microbial production and application of sophorolipids, Appl. Microbiol. Biotechnol. 76 (2007) 23–34. https://doi.org/10.1007/s00253-007-0988-7.

[37] A. Koh, R.J. Linhardt, R. Gross, Effect of Sophorolipid n-Alkyl Ester Chain Length on Its Interfacial Properties at the Almond Oil-Water Interface, Langmuir. 32 (2016) 5562–5572. https://doi.org/10.1021/acs.langmuir.6b01008.

[38] A. Koh, R. Gross, A versatile family of sophorolipid esters: Engineering surfactant structure for stabilization of lemon oil-water interfaces, Colloids Surfaces A Physicochem. Eng. Asp. 507 (2016) 152–163. https://doi.org/10.1016/j.colsurfa.2016.07.089.

[39] A. Koh, R. Gross, Molecular editing of sophorolipids by esterification of lipid moieties: Effects on interfacial properties at paraffin and synthetic crude oil-water interfaces, Colloids Surfaces A Physicochem. Eng. Asp. 507 (2016) 170–181. https://doi.org/10.1016/j.colsurfa.2016.07.084.

[40] A. Koh, A. Wong, A. Quinteros, C. Desplat, R. Gross, Influence of Sophorolipid Structure on Interfacial Properties of Aqueous-Arabian Light Crude and Related Constituent Emulsions, JAOCS, J. Am. Oil Chem. Soc. 94 (2017) 107–119. https://doi.org/10.1007/s11746-016-2913-7.

[41] N. Baccile, J.S. Pedersen, G. Pehau-Arnaudet, I.N. a. Van Bogaert, Surface charge of acidic sophorolipid micelles: effect of base and time, Soft Matter. 9 (2013) 4911–4922. https://doi.org/10.1039/c3sm50160d.

[42] G. Ben Messaoud, L. Promeneur, M. Brennich, S.L.K.W. Roelants, P. Le Griel, N. Baccile, Complex coacervation of natural sophorolipid bolaamphiphile micelles with cationic polyelectrolytes, Green Chem. 20 (2018) 3371–3385. https://doi.org/10.1039/c8gc01531g.

[43] N. Baccile, A.S. Cuvier, C. Valotteau, I.N.A. Van Bogaert, Practical methods to reduce impurities for gram-scale amounts of acidic sophorolipid biosurfactants, Eur. J. Lipid Sci. Technol. 115 (2013) 1404–1412. https://doi.org/10.1002/ejlt.201300131.

[44] P. Dhasaiyan, P. Le Griel, S. Roelants, E. Redant, I.N.A. Van Bogaert, S. Prevost, B.L.V. Prasad, N. Baccile, Micelles versus Ribbons: How Congeners Drive the Self-Assembly of Acidic Sophorolipid Biosurfactants, ChemPhysChem. 18 (2017) 643–652. https://doi.org/10.1002/cphc.201601323.

[45] J. Schindelin, I. Arganda-Carreras, E. Frise, V. Kaynig, M. Longair, T. Pietzsch, S. Preibisch, C. Rueden, S. Saalfeld, B. Schmid, J.Y. Tinevez, D.J. White, V. Hartenstein, K. Eliceiri, P. Tomancak, A. Cardona, Fiji: An open-source platform for biological-image analysis, Nat. Methods. 9 (2012) 676–682. https://doi.org/10.1038/nmeth.2019.

[46] L. Payet, E.M. Terentjev, Emulsification and stabilization mechanisms of O/W emulsions in the presence of chitosan, Langmuir. 24 (2008) 12247–12252. https://doi.org/10.1021/la8019217.

[47] X.Y. Wang, M.C. Heuzey, Chitosan-Based Conventional and Pickering Emulsions with Long-Term Stability, Langmuir. 32 (2016) 929–936. https://doi.org/10.1021/acs.langmuir.5b03556.

[48] X. Ren, Y. Zhang, Switching Pickering emulsion stabilized by Chitosan-SDS complexes through ion competition, Colloids Surfaces A Physicochem. Eng. Asp. 587 (2020) 124316. https://doi.org/10.1016/j.colsurfa.2019.124316.

[49] T.D.A. Senra, S.P. Campana-Filho, J. Desbrières, Surfactant-polysaccharide complexes

[49] based on quaternized chitosan. Characterization and application to emulsion stability, Eur. Polym. J. 104 (2018) 128–135. https://doi.org/10.1016/j.eurpolymj.2018.05.002.

[50] X. Ren, S. He, D. Liu, Y. Zhang, Multistimuli-Responsive Pickering Emulsion Stabilized by Se-Containing Surfactant-Modified Chitosan, J. Agric. Food Chem. 68 (2020) 3986–3994. https://doi.org/10.1021/acs.jafc.0c00010.

[51] N. Baccile, F. Babonneau, J. Jestin, G. Pehau-Arnaudet, I. Van Bogaert, Unusual, pH-induced, self-assembly of sophorolipid biosurfactants, ACS Nano. 6 (2012) 4763–4776. https://doi.org/10.1021/nn204911k.

[52] P. Finkle, H.D. Draper, J.H. Hildebrand, The Theory of Emulsification, J. Am. Chem. Soc. 45 (1923) 2780–2788. https://doi.org/10.1021/ja01665a002.

[53] K.A. Geisel, Microgels at Oil-Water Interfaces : Deformation , Assembly and Compression, (2015) 177.

[54] N.R. Cameron, D.C. Sherrington, High Internal Phase Emulsions (HIPEs) - Structure, Properties and Use in Polymer Preparation, Adv. Polym. Sci. 126 (1996) 162–214. https://doi.org/10.1007/3-540-60484-7_4.

[55] M. Tebboth, A. Menner, A. Kogelbauer, A. Bismarck, Polymerised high internal phase emulsions for fluid separation applications, Curr. Opin. Chem. Eng. 4 (2014) 114–120. https://doi.org/10.1016/j.coche.2014.03.001.

[56] S. Liu, M. Jin, Y. Chen, H. Gao, X. Shi, W. Cheng, L. Ren, Y. Wang, High internal phase emulsions stabilised by supramolecular cellulose nanocrystals and their application as cell-adhesive macroporous hydrogel monoliths, J. Mater. Chem. B. 5 (2017) 2671–2678. https://doi.org/10.1039/c7tb00145b.

[57] A. Sharkawy, M.F. Barreiro, A.E. Rodrigues, Chitosan-based Pickering emulsions and their applications: A review, Carbohydr. Polym. 250 (2020) 116885. https://doi.org/10.1016/j.carbpol.2020.116885.

[58] T.G. Mason, J. Bibette, D.A. Weitz, Elasticity of compressed emulsions, Phys. Rev. Lett. 75 (1995) 2051–2054. https://doi.org/10.1103/PhysRevLett.75.2051.

[59] H.M. Princen, A.D. Kiss, Rheology of foams and highly concentrated emulsions. III. Static shear modulus, J. Colloid Interface Sci. 112 (1986) 427–437. https://doi.org/10.1016/0021-9797(86)90111-6.

[60] R. Foudazi, I. Masalova, A.Y. Malkin, The rheology of binary mixtures of highly concentrated emulsions: Effect of droplet size ratio, J. Rheol. (N. Y. N. Y). 56 (2012) 1299. https://doi.org/10.1122/1.4736556.

[61] N. Romero, A. Cárdenas, M. Henríquez, H. Rivas, Viscoelastic properties and stability of highly concentrated bitumen in water emulsions, Colloids Surfaces A Physicochem. Eng. Asp. 204 (2002) 271–284. https://doi.org/10.1016/S0927-7757(02)00018-3.

[62] M. Anvari, H.S. Joyner (Melito), Effect of formulation on structure-function relationships of concentrated emulsions: Rheological, tribological, and microstructural characterization, Food Hydrocoll. 72 (2017) 11–26. https://doi.org/10.1016/j.foodhyd.2017.04.034.

[63] R.E. Guerra, Elasticity of Compressed Emulsions A dissertation presented, 2014.

[64] X. Song, Y. Pei, M. Qiao, F. Ma, H. Ren, Q. Zhao, Preparation and characterizations of Pickering emulsions stabilized by hydrophobic starch particles, Food Hydrocoll. 45 (2015) 256–263. https://doi.org/10.1016/j.foodhyd.2014.12.007.

[65] J. Xiao, X. Wang, A.J. Perez Gonzalez, Q. Huang, Kafirin nanoparticles-stabilized Pickering emulsions: Microstructure and rheological behavior, Food Hydrocoll. 54 (2016) 30–39. https://doi.org/10.1016/j.foodhyd.2015.09.008.

[66] L. Bai, S. Huan, W. Xiang, O.J. Rojas, Pickering emulsions by combining cellulose nanofibrils and nanocrystals: Phase behavior and depletion stabilization, Green Chem. 20 (2018) 1571–1582. https://doi.org/10.1039/c8gc00134k.

[67] Z. Li, T. Ngai, Stimuli-responsive gel emulsions stabilized by microgel particles, Colloid Polym. Sci. 289 (2011) 489–496. https://doi.org/10.1007/s00396-010-2362-z.

[68] J.C. Knudsen, L.H. Øgendal, L.H. Skibsted, Droplet surface properties and rheology of concentrated oil in water emulsions stabilized by heat-modified β-lactoglobulin B, Langmuir. 24 (2008) 2603–2610. https://doi.org/10.1021/la703810g.

[69] E. Guzmán, S. Llamas, A. Maestro, L. Fernández-Peña, A. Akanno, R. Miller, F. Ortega, R.G. Rubio, Polymer-surfactant systems in bulk and at fluid interfaces, Adv. Colloid Interface Sci. 233 (2016) 38–64. https://doi.org/10.1016/j.cis.2015.11.001.

[70] A. Asnacios, D. Langevin, J.F. Argillier, Mixed monolayers of cationic surfactants and anionic polymers at the air-water interface: Surface tension and ellipsometry studies, 1998.

[71] A. Asnacios, R. Klitzing, D. Langevin, Mixed monolayers of polyelectrolytes and surfactants at the air-water interface, Colloids Surfaces A Physicochem. Eng. Asp. 167 (2000) 189–197. https://doi.org/10.1016/S0927-7757(99)00475-6.

[72] D.J.F. Taylor, R.K. Thomas, J.D. Hines, K. Humphreys, J. Penfold, The adsorption of oppositely charged polyelectrolyte/surfactant mixtures at the air/water interface: Neutron reflection from dodecyl trimethylammonium bromide/sodium poly(styrene sulfonate) and sodium dodecyl sulfate/poly(vinyl pyridinium chloride), Langmuir. 18 (2002) 9783–9791. https://doi.org/10.1021/la020503d.

[73] C. Monteux, C.E. Williams, J. Meunier, O. Anthony, V. Bergeron, Adsorption of Oppositely Charged Polyelectrolyte/Surfactant Complexes at the Air/Water Interface: Formation of Interfacial Gels, Langmuir. 20 (2004) 57–63. https://doi.org/10.1021/la0347861.

[74] E.D. Goddard, Polymer/surfactant interaction: Interfacial aspects, J. Colloid Interface Sci. 256 (2002) 228–235. https://doi.org/10.1006/jcis.2001.8066.

[75] A. Klebanau, N. Kliabanova, F. Ortega, F. Monroy, R.G. Rubio, V. Starov, Equilibrium behavior and dilational rheology of polyelectrolyte/insoluble surfactant adsorption films: Didodecyldimethylammonium bromide and sodium poly(styrenesulfonate), J. Phys. Chem. B. 109 (2005) 18316–18323. https://doi.org/10.1021/jp051862v.

[76] R.A. Campbell, P.A. Ash, C.D. Bain, Dynamics of adsorption of an oppositely charged polymer-surfactant mixture at the air-water interface: Poly(dimethyldiallylamnionium chloride) and sodium dodecyl sulfate, Langmuir. 23 (2007) 3242–3253. https://doi.org/10.1021/la0632171.

[77] A. Angus-Smyth, C.D. Bain, I. Varga, R.A. Campbell, Effects of bulk aggregation on PEI-SDS monolayers at the dynamic air-liquid interface: Depletion due to precipitation versus enrichment by a convection/spreading mechanism, Soft Matter. 9 (2013) 6103–6117. https://doi.org/10.1039/c3sm50636c.

[78] M. Destribats, V. Lapeyre, E. Sellier, F. Leal-Calderon, V. Schmitt, V. Ravaine, Water-in-oil emulsions stabilized by water-dispersible poly(N- isopropylacrylamide) microgels: Understanding anti-Finkle behavior, Langmuir. 27 (2011) 14096–14107. https://doi.org/10.1021/la203476h.

[79] D. Kurukji, R. Pichot, F. Spyropoulos, I.T. Norton, Interfacial behaviour of sodium stearoyllactylate (SSL) as an oil-in-water pickering emulsion stabiliser, J. Colloid Interface Sci. 409 (2013) 88–97. https://doi.org/10.1016/j.jcis.2013.07.016.

[80] M. Destribats, M. Rouvet, C. Gehin-Delval, C. Schmitt, B.P. Binks, Emulsions stabilised by whey protein microgel particles: Towards food-grade Pickering emulsions, Soft Matter. 10 (2014) 6941–6954. https://doi.org/10.1039/c4sm00179f.

[81] G. Dardelle, M. Jacquemond, P. Erni, Delivery Systems for Low Molecular Weight Payloads: Core/Shell Capsules with Composite Coacervate/Polyurea Membranes, Adv.

Mater. 29 (2017) 1606099. https://doi.org/10.1002/adma.201606099.
[82] A.M. Bago Rodriguez, B.P. Binks, T. Sekine, Novel stabilisation of emulsions by soft particles: Polyelectrolyte complexes, Faraday Discuss. 191 (2016) 255–285. https://doi.org/10.1039/c6fd00011h.
[83] A.M. Bago Rodriguez, B.P. Binks, T. Sekine, Emulsions Stabilized with Polyelectrolyte Complexes Prepared from a Mixture of a Weak and a Strong Polyelectrolyte, Langmuir. 35 (2019) 6693–6707. https://doi.org/10.1021/acs.langmuir.9b00897.
[84] B.P. Binks, Chapter 1. Emulsions — Recent Advances in Understanding, in: Mod. Asp. Emuls. Sci., Royal Society of Chemistry, 2007: pp. 1–55. https://doi.org/10.1039/9781847551474-00001.
[85] Z. Hu, S. Ballinger, R. Pelton, E.D. Cranston, Surfactant-enhanced cellulose nanocrystal Pickering emulsions, J. Colloid Interface Sci. 439 (2015) 139–148. https://doi.org/10.1016/j.jcis.2014.10.034.
[86] S. Huan, S. Yokota, L. Bai, M. Ago, M. Borghei, T. Kondo, O.J. Rojas, Formulation and Composition Effects in Phase Transitions of Emulsions Costabilized by Cellulose Nanofibrils and an Ionic Surfactant, Biomacromolecules. 18 (2017) 4393–4404. https://doi.org/10.1021/acs.biomac.7b01452.
[87] L. Bai, W. Xiang, S. Huan, O.J. Rojas, Formulation and Stabilization of Concentrated Edible Oil-in-Water Emulsions Based on Electrostatic Complexes of a Food-Grade Cationic Surfactant (Ethyl Lauroyl Arginate) and Cellulose Nanocrystals, Biomacromolecules. 19 (2018) 1674–1685. https://doi.org/10.1021/acs.biomac.8b00233.

# Supporting Information

## pH-Switchable Pickering Emulsions Stabilized by Biosurfactant-Polyelectrolyte Complex Coacervate Colloids

Sandrine Laquerbe,[a] Alain Carvalho,[b] Marc Schmutz,[b] Alexandre Poirier,[a] Niki Baccile,[a,*] Ghazi Ben Messaoud[a,†,*]

[a] Sorbonne Université, Centre National de la Recherche Scientifique, Laboratoire de Chimie de la Matière Condensée de Paris, LCMCP, F-75005 Paris, France

[b] Université de Strasbourg, CNRS, Institut Charles Sadron UPR 22, 67034 Strasbourg, France

† Current address: DWI- Leibniz Institute for Interactive Materials, Forckenbeckstrasse 50, 52056, Aachen, Germany

**Content:**

Figure S1 to Figure S6

Video 1 to Video 3, available free of charge on the Journal's website.

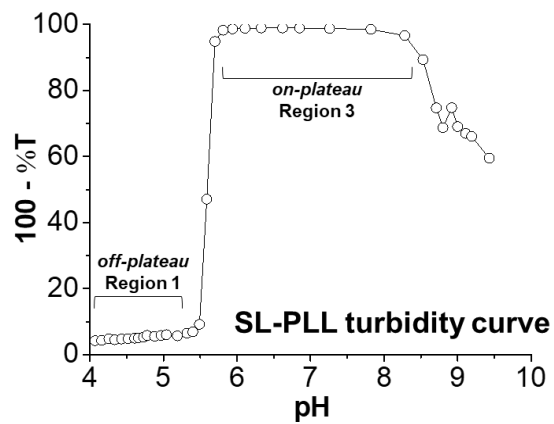

**Figure S 1 - Turbidity profile (100%-T) as a function of pH for SL-PLL mixture ($C_{SL}$= 5 mg/mL; $C_{PLL}$= 2 mg/mL, data reproduced from Ben Messaoud et al.,[1] Complex coacervation is maximized *on-plateau*, between pH 6 and pH 8.5**

---

[1] G. Ben Messaoud, L. Promeneur, M. Brennich, S.L. Roelants, P. Le Griel and N. Baccile (2018). Complex coacervation of natural sophorolipid bolaamphiphile micelles with cationic polyelectrolytes. *Green Chemistry*, *20*(14), 3371-3385.

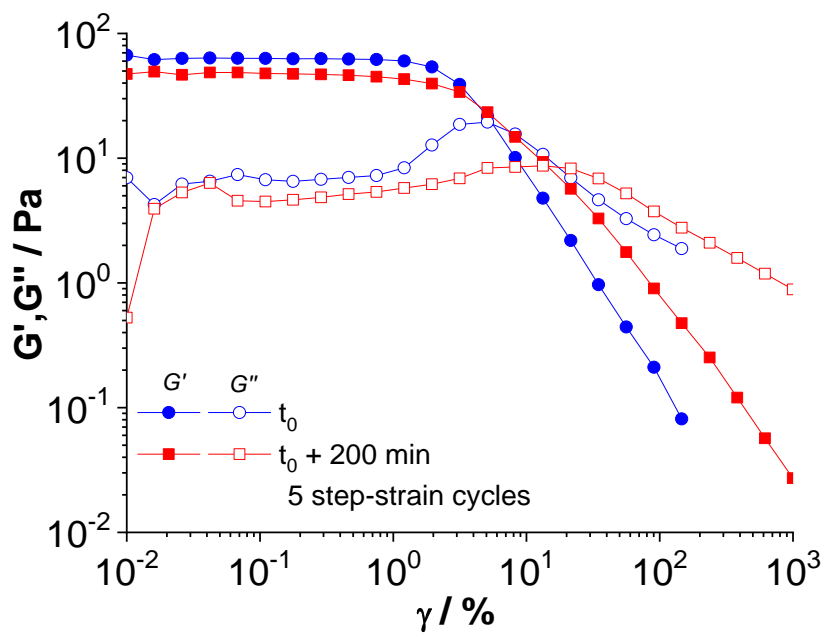

**Figure S 2 -** Comparison of storage (G') and loss (G'') moduli as a function of strain for o/w emulsion gels ($\Phi_{oil}$= 0.7) composed of SL-PLL complex coacervates ($C_{SL}$= 5 mg/mL; $C_{PLL}$= 2 mg/mL) before ($t_0$) and after ($t_0$ + 200 min) five step-strain ($\omega$= 6.28 rad/s) cycles presented in Figure 6 in the main text.

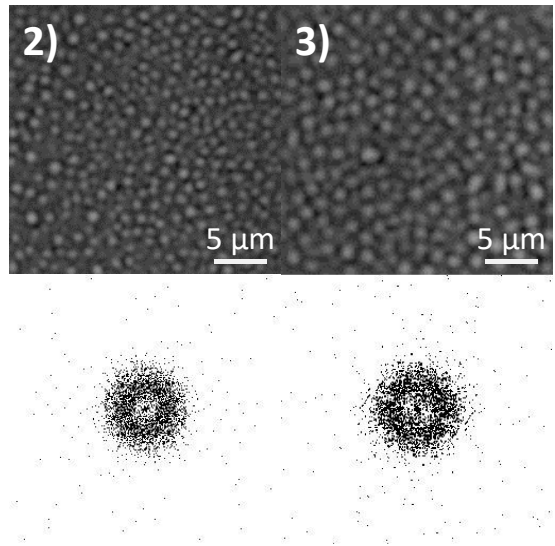

**Figure S 3 – Magnification of regions (2) and (3) and related Fourier Transform corresponding to the Single labeling experiment shown in Figure 8b in the main text.**

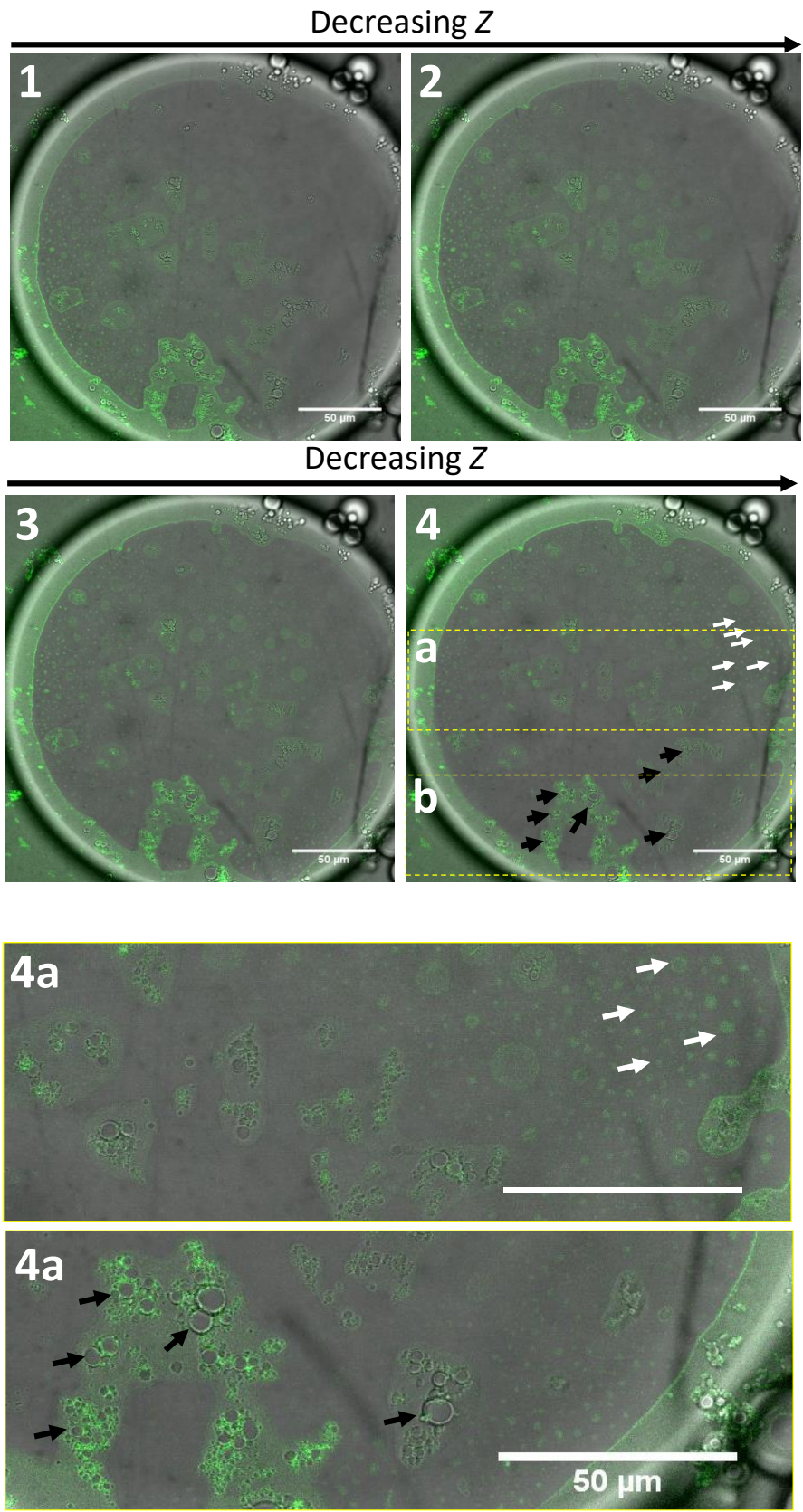

**Figure S 4 – CLSM images from single labeling experiments of o/w emulsion at pH= 6.27. Merged white and PLL-FITC green channel of a typical gelled emulsion composed of water and SL-PLL complex coacervates prepared at a plateau pH value. Green particles correspond to complex coacervates**

containing FITC-labeled PLL. These images focus on the coexistence of oil droplets and complex coacervates at the surface of a large oil droplet. Black arrow point at core-shell spheres, of which only the shell is fluorescent, while the interior is fluorophore-free. These objects are attributed to oil droplets adsorbed on the surface of the large droplet. White arrows point at dense green colloids of the same nature as the ones highlighted in Figure 8c of the main text and Figure S 3 in the Supporting Information. "Decreasing *Z*" indicates the descending direction of the focal planes, from 1 to 4 (plane 1 being higher than plane 4), in the *Z* (height) direction of the confocal microscope.

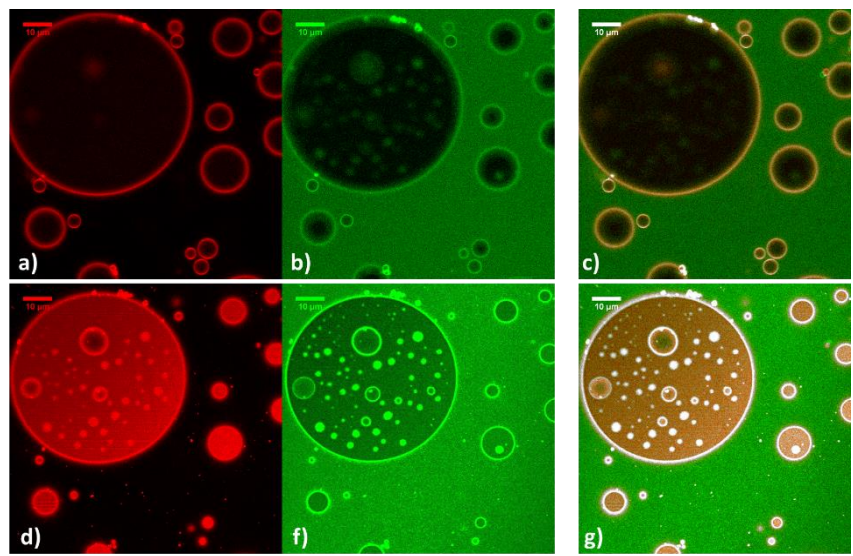

**Figure S 5 – Additional images of a *Double labeling experiment*:** a,d) 583 nm (red), b,f) 520 nm (green) and c,g) merged (red, green) channel CLSM images of a typical gelled emulsion composed of water and SL-PLL complex coacervates. In c) and f), the white/yellowish region/layers are obtained through the *colocalization* function of Fiji: only regions where red and green channels are colocalized are shown in white. For all experiments, the selected z-plane corresponds to the top surface of the oil droplets. Details on methodology for each experiment and composition are given in Table 1 in the materials and methods section of the main text.

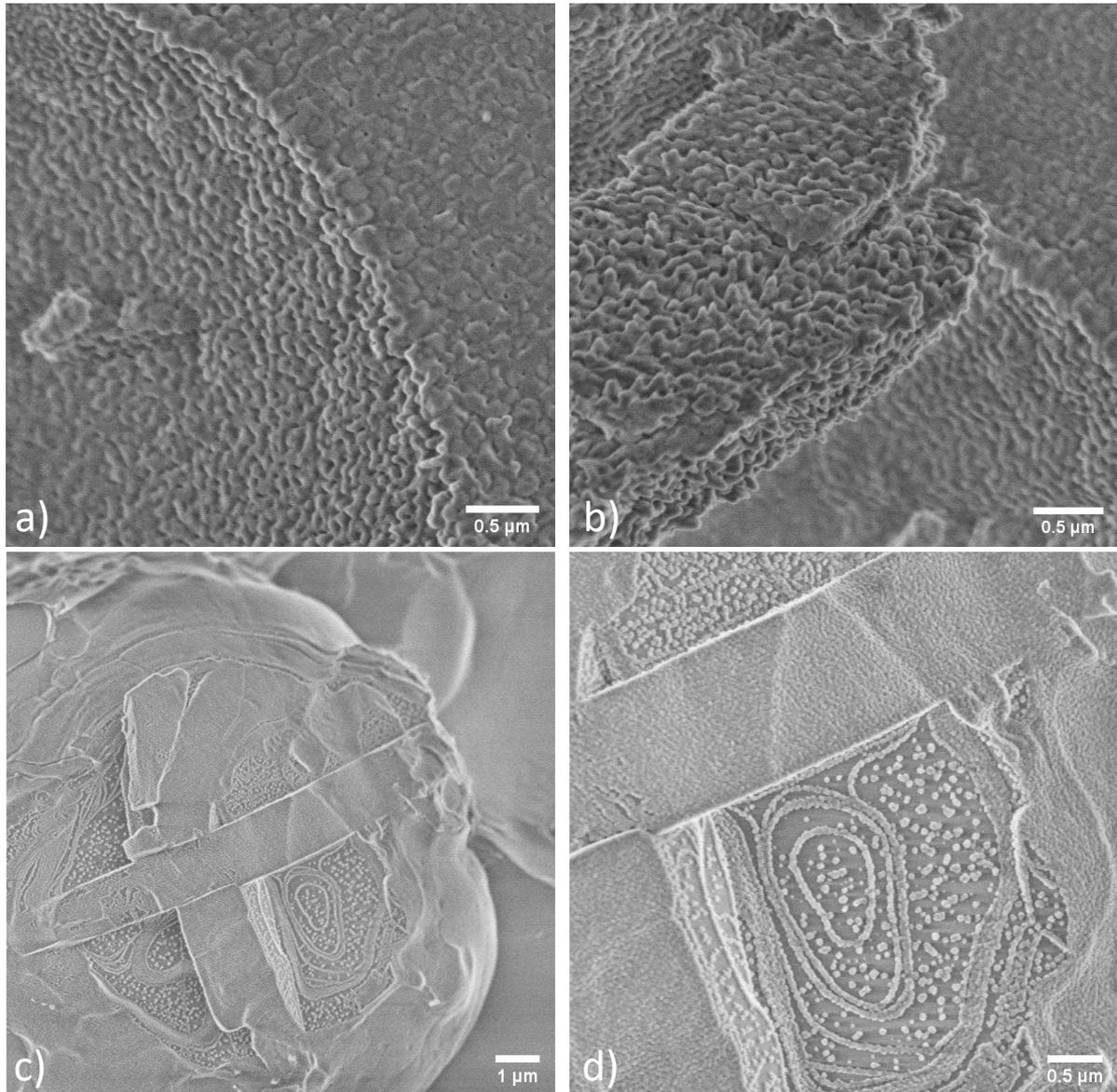

**Figure S 6** – Additional cryo-SEM images of o/w emulsions (Φoil= 0.7) and an aqueous solution of complex coacervates composed of a-b) SL and PLL ($C_{SL}$= 5 mg/mL; $C_{PLL}$= 2 mg/mL) at pH= 6.55. c-d) SL and CHL ($C_{SL}$= 5 mg/mL; $C_{CHL}$= 1.4 mg/mL) at pH= 6.22.